\documentclass[onecolumn,authoryear]{elsarticle}

\usepackage[a4paper,margin=2.5cm]{geometry}
\usepackage{graphicx}      
\usepackage{natbib}        
\usepackage[dvipsnames]{xcolor}
\usepackage{amssymb}
\usepackage{amsmath}

\usepackage{tikz}
\usepackage{bm}
\usepackage{siunitx}
\usepackage{tabularx}
\usepackage{booktabs}
\usepackage{hyperref}
\hypersetup{
	colorlinks,
	linkcolor={MidnightBlue},
	citecolor={MidnightBlue},
	urlcolor={MidnightBlue}
} 

\newtheorem{theorem}{\textbf{Theorem}}

\newtheorem{remark}{\textbf{Remark}}

\newcommand{\bbx}{\mathbf{x}}
\newcommand{\bbu}{\mathbf{u}}
\newcommand{\bbw}{\mathbf{w}}
\newcommand{\bbup}{\bbu_{\mathrm{p}}}

\newcommand{\cN}{\mathcal{N}}
\newcommand{\cU}{\mathcal{U}}
\newcommand{\cI}{\mathcal{I}}
\newcommand{\cD}{\mathcal{D}}
\newcommand{\cL}{\mathcal{L}}
\newcommand{\cX}{\mathcal{X}}
\newcommand{\cW}{\mathcal{W}}

\newcommand{\bN}{\mathbb{N}}
\newcommand{\bR}{\mathbb{R}}
\newcommand{\bI}{\mathbb{I}}

\newcommand{\bw}{\mathbf{w}}

\newcommand{\Ts}{T_{\mathrm{s}}}
\newcommand{\Tsim}{T_{\mathrm{sim}}}
\newcommand{\Nsim}{N_{\mathrm{sim}}}
\newcommand{\bzero}{\mathbf{0}}
\newcommand{\bone}{\mathbf{1}}

\newcommand{\wbp}{w^+_{\beta}}
\newcommand{\wbm}{w^-_{\beta}}
\newcommand{\wbb}{w^\ast_{\beta}(t,\beta(t))}

\newcommand{\Nfg}{N_{\mathrm{fg}}}

\newcommand{\dfg}[1]{\delta_{\mathrm{fg},#1}}

\newcommand{\Pfg}[1]{P_{\mathrm{fg},#1}}
\newcommand{\Pfgsys}{P^+_{\mathrm{fg}}}

\newcommand{\bPfg}{\mathbf{P}_{\mathrm{fg}}}
\newcommand{\Cfg}[1]{C_{\mathrm{fg},#1}}
\newcommand{\Ofg}[1]{O_{\mathrm{fg,#1}}}
\newcommand{\Sfg}[1]{S_{\mathrm{fg},#1}}
\newcommand{\zfg}[1]{z_{\mathrm{fg},#1}}

\newcommand{\Ness}{M_{\mathrm{ess}}}
\newcommand{\Pess}{P_{\mathrm{ess}}}
\newcommand{\bPess}{\bar{P}_{\mathrm{ess}}}
\newcommand{\xess}{x_{\mathrm{ess}}}
\newcommand{\xesst}{x^{\ast}_{\mathrm{ess}}}

\newcommand{\dess}{\delta_{\mathrm{ess}}}
\newcommand{\Cess}{C_{\mathrm{ess}}}
\newcommand{\Oess}{O_{\mathrm{ess}}}
\newcommand{\etac}{\eta_{\mathrm{c}}}
\newcommand{\etad}{\eta_{\mathrm{d}}}
\newcommand{\zess}{z_{\mathrm{ess}}}
\newcommand{\xdg}{x_{\mathrm{dg}}}
\newcommand{\fess}{f_{\mathrm{ess}}}

\newcommand{\Pres}{P_{\mathrm{res}}}
\newcommand{\Prest}{P^{\ast}_{\mathrm{res}}}
\newcommand{\hPres}{\widehat{P}_{\mathrm{res}}}

\newcommand{\Pload}{P_{\mathrm{load}}}
\newcommand{\Ploadt}{P^{\ast}_{\mathrm{load}}}
\newcommand{\hPload}{\widehat{P}_{\mathrm{load}}}
\newcommand{\Cload}{C_{\mathrm{load}}}

\newcommand{\Pexg}{P_{\mathrm{exg}}}
\newcommand{\bPexg}{\bar{P}_{\mathrm{exg}}}
\newcommand{\Cexg}{C_{\mathrm{exg}}}
\newcommand{\dexg}{\delta_{\mathrm{exg}}}
\newcommand{\cs}{c_{\mathrm{s}}}
\newcommand{\cp}{c_{\mathrm{p}}}
\newcommand{\zexg}{z_{\mathrm{exg}}}

\newcommand{\Cgrid}{C_{\mathrm{grid}}}

\newcommand{\pmpc}{\mathrm{P}_{\mathrm{MPC}}}
\newcommand{\umpct}[1]{\tilde{u}_{\mathrm{MPC},#1}}
\newcommand{\mmpct}[1]{\tilde{\nu}_{\mathrm{MPC},#1}}
\newcommand{\pimpc}{\pi_{\mathrm{MPC}}}
\newcommand{\pimlp}[1]{\pi_{\mathrm{MLP,#1}}}
\newcommand{\Proj}{\text{Proj}}

\newcommand{\qt}[1]{{\color{black}#1}}
\newcommand{\kk}[1]{{\color{black}#1}}

\journal{Engineering Applications in Artificial Intelligence}

\begin{document}

\begin{frontmatter}
	
	
	
	\title{Approximate Model Predictive Control for Microgrid Energy Management via Imitation Learning} 
	\tnotetext[t1]{This paper is part of a project that has received funding from the European Research Council (ERC) under the European Union’s Horizon 2020 research and innovation programme (Grant agreement No. 101018826 - CLariNet) and US NSF ECCS No. 2234032.}
	
	
	\author[1]{Changrui Liu\corref{cor1}}
	\ead{c.liu-14@tudelft.nl}
	
	\author[1]{Shengling Shi}
	\ead{shengling.shi@tudelft.nl}
	\author[1]{Anil Alan}
	\ead{a.alan@tudelft.nl}
	
	\author[2,3]{Ganesh Kumar Venayagamoorthy}
	\ead{gkumar@ieee.org}
	
	\author[1]{Bart De Schutter}
	\ead{b.deschutter@tudelft.nl}
	
	\cortext[cor1]{Corresponding author}
	
	\affiliation[1]{organization={Delft Center for System and Control, TU Delft},
		addressline={Mekelweg 2}, 
		city={Delft},
		postcode={2628CD}, 
		state={},
		country={The Netherlands}}
		
	\affiliation[2]{organization={Real-Time Power and Intelligent Systems Laboratory, Department of Electrical and Computer Engineering, Clemson Universit},
		addressline={105 Sikes Hall}, 
		city={Clemson},
		postcode={SC 29634}, 
		state={South Carolina},
		country={U.S.}}
		
	\affiliation[3]{organization={Department of Electrical, Electronic and Computer Engineering, University of Pretoria},
		addressline={Lynnwood Road, Hatfield}, 
		city={Pretoria},
		postcode={0028}, 
		state={South Carolina},
		country={South Africa}}	
	\begin{abstract}
		Efficient energy management is essential for reliable and sustainable microgrid operation amid increasing renewable integration. In this paper, an imitation learning–based framework to approximate mixed-integer Economic Model Predictive Control (EMPC) is proposed for microgrid energy management, \qt{considering fuel generators, renewable energy resources, a unified energy storage unit, and curtailable loads.} Within the proposed framework, a neural network is trained to imitate expert EMPC control actions from offline trajectories, thereby enabling fast real-time decision making without solving online \qt{mixed-integer} optimization problems, \qt{which often exhibit highly variable solution times across instances and do not scale well to large problem sizes; in particular, worst-case solve times can be excessively large and therefore unsuitable for real-time deployment. In contrast, the learned policy provides predictable and consistently low computation times.} To enhance robustness and generalization, the learning process incorporates noise injection during training to mitigate distribution shift and explicitly accounts for forecast uncertainty in renewable generation and demand. \kk{Furthermore, a constraint-tightening approach combined with a projection layer is proposed to ensure recursive feasibility and constraint satisfaction of the learned controller.} Simulation results demonstrate that the learned policy achieves economic performance comparable to EMPC, \kk{while reducing computation time by approximately one order of magnitude relative to the optimization-based EMPC}.
	\end{abstract}
	
	%
	
	
\end{frontmatter}


\section{Introduction}
\vspace*{-0.2cm}
\label{sec:1-intro}
The integration of distributed energy resources, including photovoltaics, wind turbines, and Energy Storage Systems (ESSs), has led to the growing deployment of microgrids, which enable more localized generation, storage, and consumption of electricity \citep{chen2020networked}. Designing an effective Energy Management System (EMS) for microgrids is crucial for ensuring reliability and reducing operational costs under uncertain generation and load profiles. The core of an EMS lies in solving an optimization problem, for which a wide range of methods have been explored, including quadratic programming~\citep{yoon2020quadratic}, stochastic programming that explicitly accounts for uncertainties in generation and demand~\citep{liang2014stochastic}, and meta-heuristic approaches that address non-convex or non-smooth problem structures~\citep{akter2024review}. The interested reader is referred to the comprehensive review by~\cite{zia2018microgrids} for additional methods in this area. Recent developments have also focused on the use of Model Predictive Control (MPC)~\qt{\citep{prodan2014model, pippia2019single, alarcon2022economic, hu2023economic, alarcon2023scenario}}, which has emerged as an appealing framework because it naturally integrates system dynamics, forecast information, and operational constraints within an optimization-based control scheme. In particular, Economic MPC (EMPC) has attracted considerable attention~\citep{parisio2014model, tobajas2022resilience, hu2023economic}, as it extends conventional MPC by directly optimizing an economic performance metric, such as operating cost or profit, instead of focusing solely on \qt{stabilization or reference tracking}. This formulation makes EMPC well suited for modern microgrids, where achieving economic efficiency and sustainable energy integration are primary objectives.

Recognizing its potential benefits, numerous studies have implemented EMPC for microgrid energy management, demonstrating its practical effectiveness~\citep{hu2023economic}. \cite{parisio2014model} incorporated controllable loads into the model, allowing load curtailment as part of the energy management strategy. \cite{tobajas2022resilience} considered a hybrid ESS, resulting in a mixed-logical dynamical model in the EMPC formulation. In general, logic variables provide a compact and effective means to represent the discrete behaviors commonly encountered in microgrids, such as the ON/OFF status of fuel generators, charging/discharging modes of ESS, and purchasing/selling decisions when exchanging electricity with the main grid \citep{pippia2019single, tobajas2022resilience, hu2023economic, dasilva2025integrating}. Consequently, when MPC is applied to such systems, mixed-integer MPC (MI-MPC) formulations~\citep{karg2018deep} are typically required to handle both continuous and discrete dynamics. However, a major limitation of MI-MPC lies in its substantial computational burden. At each sampling instant, a constrained mixed-integer program must be solved online, which can become prohibitively expensive for large-scale or fast-evolving microgrids~\citep{pippia2019single, gao2021online, dasilva2025integrating}. This challenge is further exacerbated by the presence of longer prediction horizons, nonlinear system dynamics, and short sampling intervals, all of which significantly hinder real-time implementation even when using advanced optimization solvers. \qt{While bi-level or hierarchical MPC formulations can alleviate part of the computational burden of hybrid MPC~\citep{guo2016optimal}, discrete optimization typically remains in the upper-level problem, meaning that mixed-integer programs must still be solved online. Consequently, the computational cost can remain significant and difficult to predict in real-time applications. In contrast, approximate MPC computes inputs via a simple policy evaluation with constant and predictable computation times.}

Following several seminal works on approximating MPC policies using machine learning~\citep{chen2018approximating, drgovna2018approximate, hertneck2018learning, karg2018deep}, collectively known as learning-based approximate MPC, extensive research has explored the use of different neural networks for approximate MPC~\citep{shen2024generative, tong2025ensemble}. In parallel, some papers have focused on improving sample efficiency through fast data augmentation~\citep{krishnamoorthy2021sensitivity} and providing safety guarantees by adding a projection layer to the network~\citep{hose2025approximate}. A key advantage of approximate MPC lies in its ability to enable fast online computation, making it particularly attractive for real-time applications. As a result, it has been successfully applied to domains such as smart buildings~\citep{drgovna2018approximate, karg2018deep} and solar trough plants \citep{ruiz2025electric}. 

\qt{In the context of microgrid energy management, machine learning techniques have been widely used, including demand prediction~\citep{alarcon2025artificial} and learning MPC policies~\citep{gao2021online, dasilva2025integrating, alarcon2025learning}.} However, most existing approaches adopt \textit{indirect} approximate MPC, in which machine learning is primarily used to predict optimized integer or binary variables within MI-MPC formulations~\citep{gao2021online, dasilva2025integrating}. Once these discrete decisions are predicted, the remaining problem reduces to a linear or quadratic program that can be solved very efficiently. \qt{Recently, a \textit{direct} approximate MPC approach based on recurrent neural networks has also been developed~\citep{alarcon2025learning}. However, the considered microgrid models do not include fuel generators or load curtailment.} In general, results on applying \textit{direct} approximate MI-MPC to energy applications remain scarce~\citep{karg2018deep, lohr2020machine, alarcon2025learning}.

On the other hand, many approximate MPC methods in the literature rely on open-loop data uniformly sampled over a grid~\citep{chen2018approximating, hertneck2018learning}. More recent research has therefore focused on using Imitation Learning (IL) to sample directly from closed-loop trajectories during training~\citep{drgovna2018approximate, karg2018deep, pfrommer2024sample, alarcon2025learning}, with the goal of improving sample efficiency and closed-loop performance. Despite the strong potential of IL to significantly reduce the computational burden associated with solving MI-MPC problems in microgrid energy management, its application to approximate MI-MPC remains largely unexplored, \qt{with the exception of the recent work by~\cite{alarcon2025learning}, where approximate MPC is combined with a real-time optimizer that determines the optimal steady state.} Moreover, the issue of distribution shift~\citep{ross2011reduction}, namely the mismatch between the state distributions encountered during training and those visited by the learned policy during deployment, has not been adequately addressed in existing MI-MPC applications~\qt{\citep{drgovna2018approximate, karg2018deep, alarcon2025learning}}.

This study proposes an IL–based framework for approximate MI-MPC tailored to microgrid energy management \qt{in day-ahead scheduling}. The considered microgrid scheduling and operation problem includes fuel generators, renewable energy sources (RESs), curtailable loads, \qt{a unified ESS unit}, and operational constraints. \qt{While stochastic MPC formulations~\citep{pozzi2025imitation} can be used to explicitly account for uncertainty in renewable generation and loads, access to reasonably accurate forecasts is assumed in this work, and a certainty-equivalence MPC scheme~\citep{meadows1995topics, liu2026certainty} is adopted, wherein disturbances are replaced by their predicted values, resulting in a simple yet effective framework for real-time microgrid energy management~\citep{hu2021model}.} The core contribution lies in applying IL to \textit{directly} approximate the EMPC control policy, thereby replacing repeated online optimization with a lightweight, learned controller. To the best of the authors' knowledge, this is the first work to leverage imitation learning for \qt{fully end-to-end} approximation of mixed-integer EMPC for microgrid energy management. The main contributions of the current paper are as follows:
\begin{itemize}
	\item \qt{A novel IL-based approach for \emph{directly} approximating MI-MPC policies is proposed.} The approach \qt{introduces feature representations tailored to load curtailment decisions in microgrid energy management} and employs a noisy expert strategy during training to mitigate distribution shift.
	
	\item \qt{Forecast uncertainties in renewable energy generation and load demand are explicitly incorporated into both the offline data generation process and the online deployment of the learned controller, thereby enhancing robustness under realistic operating conditions.} \kk{Furthermore, a novel input constraint-tightening approach is proposed and integrated with the nominal EMPC framework. Sufficient conditions for recursive feasibility, without relying on terminal ingredients, are derived, and recursive feasibility under the proposed scheme is rigorously proven. The resulting tightened input constraints can be seamlessly embedded into the learned controller via a projection layer, ensuring satisfaction of both state and input constraints.}
	
	\item Simulation results demonstrate that the proposed approach achieves economic performance comparable to optimization-based EMPC while \qt{reducing computation time by an order of magnitude, highlighting its potential for real-time microgrid energy management.}
\end{itemize}

\qt{It is noted that this work employs a standard Multi-Layer Perceptron (MLP) as the learned controller and does not focus on exploring alternative neural network architectures or developing new network structures for approximate EMPC.} The remainder of the paper is organized as follows. Section~\ref{sec:2-modeling} introduces the microgrid model, including the logic relations. Section~\ref{sec:3-mpc} formulates the mixed-integer EMPC problem. The proposed IL-based approximate EMPC is presented in Section~\ref{sec:4-imitation}. Section~\ref{sec:5-simulation} details the simulation setup and provides comparative results. Finally, Section~\ref{sec:6-conclusion} concludes the paper and outlines directions for future research.
\begin{figure}[h]
	\centering

	\tikzset{every picture/.style={line width=0.75pt}} 
	\resizebox{0.5\columnwidth}{!}{%

	\tikzset{every picture/.style={line width=0.75pt}} 
	
	\begin{tikzpicture}[x=0.75pt,y=0.75pt,yscale=-1,xscale=1]
		
		\draw (510.13,259.77) node  {\includegraphics[width=34.56pt,height=38.66pt]{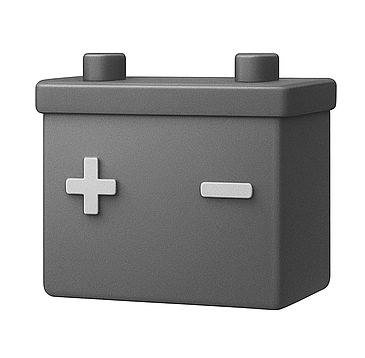}};
		\draw (298.91,151.36) node  {\includegraphics[width=31.36pt,height=35.05pt]{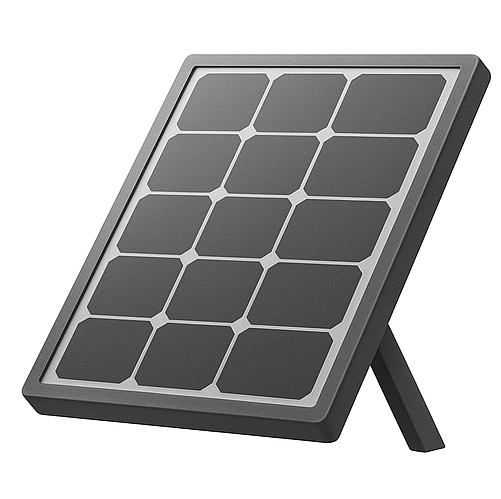}};
		\draw (340.91,150.56) node  {\includegraphics[width=29.86pt,height=46.23pt]{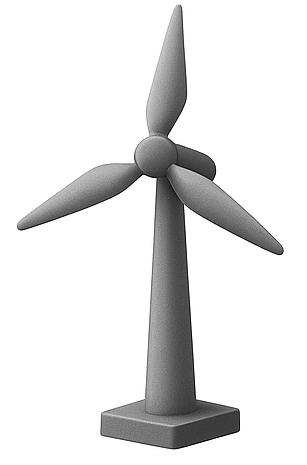}};
		\draw  [fill={rgb, 255:red, 126; green, 211; blue, 33 }  ,fill opacity=0.2 ] (273.82,133.55) .. controls (273.82,124.41) and (281.23,117) .. (290.36,117) -- (349.27,117) .. controls (358.41,117) and (365.82,124.41) .. (365.82,133.55) -- (365.82,183.18) .. controls (365.82,192.32) and (358.41,199.73) .. (349.27,199.73) -- (290.36,199.73) .. controls (281.23,199.73) and (273.82,192.32) .. (273.82,183.18) -- cycle ;
		
		\draw (329.41,261.36) node  {\includegraphics[width=38.11pt,height=36.55pt]{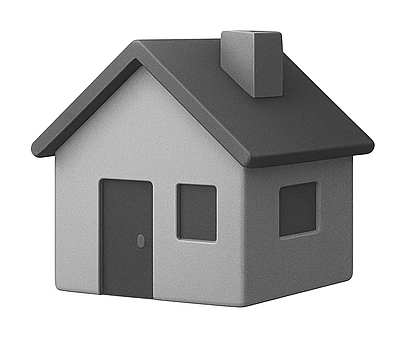}};
		\draw (380.23,261.36) node  {\includegraphics[width=38.11pt,height=36.55pt]{figure/load.png}};
		\draw  [fill={rgb, 255:red, 208; green, 2; blue, 27 }  ,fill opacity=0.2 ] (296,249.95) .. controls (296,242.8) and (301.8,237) .. (308.95,237) -- (401.87,237) .. controls (409.02,237) and (414.82,242.8) .. (414.82,249.95) -- (414.82,288.78) .. controls (414.82,295.93) and (409.02,301.73) .. (401.87,301.73) -- (308.95,301.73) .. controls (301.8,301.73) and (296,295.93) .. (296,288.78) -- cycle ;
		
		\draw (64,271) node  {\includegraphics[width=52.5pt,height=52.5pt]{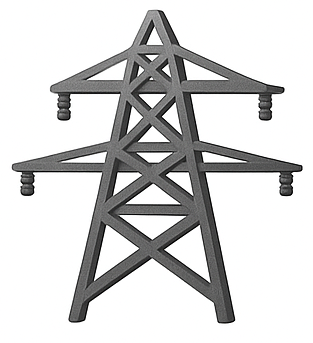}};
		\draw  [fill={rgb, 255:red, 248; green, 231; blue, 28 }  ,fill opacity=0.2 ] (21,248.89) .. controls (21,239.41) and (28.68,231.73) .. (38.16,231.73) -- (89.65,231.73) .. controls (99.13,231.73) and (106.82,239.41) .. (106.82,248.89) -- (106.82,314.56) .. controls (106.82,324.04) and (99.13,331.73) .. (89.65,331.73) -- (38.16,331.73) .. controls (28.68,331.73) and (21,324.04) .. (21,314.56) -- cycle ;
		
		\draw (441.7,149) node  {\includegraphics[width=32.56pt,height=33pt]{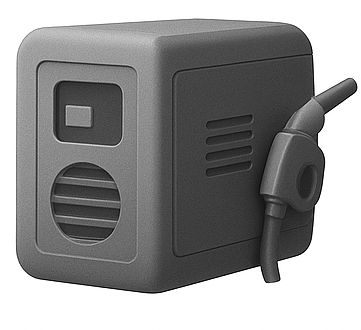}};
		\draw (485.11,149) node  {\includegraphics[width=32.56pt,height=33pt]{figure/FG.png}};
		\draw  [fill={rgb, 255:red, 126; green, 211; blue, 33 }  ,fill opacity=0.2 ] (397,136.93) .. controls (397,129.24) and (403.24,123) .. (410.93,123) -- (510.89,123) .. controls (518.58,123) and (524.82,129.24) .. (524.82,136.93) -- (524.82,178.71) .. controls (524.82,186.4) and (518.58,192.64) .. (510.89,192.64) -- (410.93,192.64) .. controls (403.24,192.64) and (397,186.4) .. (397,178.71) -- cycle ;
		
		\draw (465.04,259.77) node  {\includegraphics[width=34.56pt,height=38.66pt]{figure/ESS.png}};
		\draw  [fill={rgb, 255:red, 245; green, 166; blue, 35 }  ,fill opacity=0.2 ] (433.09,249.91) .. controls (433.09,242.78) and (438.87,237) .. (445.99,237) -- (531.18,237) .. controls (538.31,237) and (544.09,242.78) .. (544.09,249.91) -- (544.09,288.63) .. controls (544.09,295.76) and (538.31,301.54) .. (531.18,301.54) -- (445.99,301.54) .. controls (438.87,301.54) and (433.09,295.76) .. (433.09,288.63) -- cycle ;
		\draw (173,160) node  {\includegraphics[width=52.5pt,height=52.5pt]{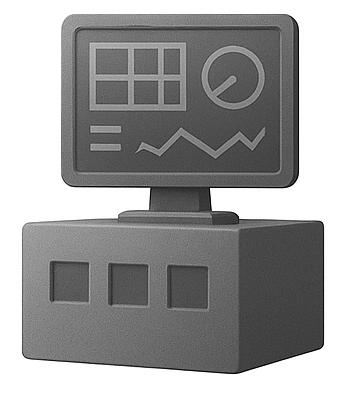}};
		\draw  [fill={rgb, 255:red, 74; green, 144; blue, 226 }  ,fill opacity=0.2 ] (133,134.76) .. controls (133,126.06) and (140.06,119) .. (148.76,119) -- (196.05,119) .. controls (204.76,119) and (211.82,126.06) .. (211.82,134.76) -- (211.82,200.15) .. controls (211.82,208.85) and (204.76,215.91) .. (196.05,215.91) -- (148.76,215.91) .. controls (140.06,215.91) and (133,208.85) .. (133,200.15) -- cycle ;
		
		\draw [line width=3]    (233.82,100) -- (232.82,351.18) ;
		\draw  [fill={rgb, 255:red, 245; green, 166; blue, 35 }  ,fill opacity=0.99 ] (107,288) -- (119.82,284) -- (119.82,286.27) -- (218,286.27) -- (218,284) -- (230.82,288) -- (218,292) -- (218,289.73) -- (119.82,289.73) -- (119.82,292) -- cycle ;
		\draw  [color={rgb, 255:red, 0; green, 0; blue, 0 }  ,draw opacity=1 ][fill={rgb, 255:red, 208; green, 2; blue, 27 }  ,fill opacity=1 ] (233.32,267.42) -- (283.09,267.42) -- (283.09,265.18) -- (294.09,269.36) -- (283.09,273.54) -- (283.09,271.3) -- (233.32,271.3) -- cycle ;
		\draw  [fill={rgb, 255:red, 184; green, 233; blue, 134 }  ,fill opacity=1 ] (236,160.55) -- (248.82,156.18) -- (248.82,158.36) -- (273.82,158.36) -- (273.82,162.73) -- (248.82,162.73) -- (248.82,164.91) -- cycle ;
		\draw  [fill={rgb, 255:red, 184; green, 233; blue, 134 }  ,fill opacity=1 ] (469.82,193.18) -- (469.83,211.63) -- (248.82,211.71) -- (248.82,210.18) -- (234.82,214.21) -- (248.82,218.24) -- (248.82,216.71) -- (474.83,216.63) -- (474.82,193.18) -- cycle ;
		\draw  [fill={rgb, 255:red, 155; green, 155; blue, 155 }  ,fill opacity=0.19 ][dash pattern={on 4.5pt off 4.5pt}][line width=0.75]  (126.1,79.36) -- (555.1,79.36) -- (555.1,380.55) -- (126.1,380.55) -- cycle ;
		\draw [color={rgb, 255:red, 74; green, 144; blue, 226 }  ,draw opacity=1 ][line width=1.5]    (171,216.36) -- (171.77,281.18) ;
		\draw [shift={(171.82,285.18)}, rotate = 269.32] [fill={rgb, 255:red, 74; green, 144; blue, 226 }  ,fill opacity=1 ][line width=0.08]  [draw opacity=0] (13.4,-6.43) -- (0,0) -- (13.4,6.44) -- (8.9,0) -- cycle    ;
		\draw [color={rgb, 255:red, 74; green, 144; blue, 226 }  ,draw opacity=1 ][line width=1.5]    (188,216.36) -- (187.82,246.36) ;
		\draw [color={rgb, 255:red, 74; green, 144; blue, 226 }  ,draw opacity=1 ][line width=1.5]    (186.82,246.36) -- (267.82,246.36) ;
		\draw [color={rgb, 255:red, 74; green, 144; blue, 226 }  ,draw opacity=1 ][line width=1.5]    (267.82,245.36) -- (267.82,262.36) ;
		\draw [shift={(267.82,266.36)}, rotate = 270] [fill={rgb, 255:red, 74; green, 144; blue, 226 }  ,fill opacity=1 ][line width=0.08]  [draw opacity=0] (13.4,-6.43) -- (0,0) -- (13.4,6.44) -- (8.9,0) -- cycle    ;
		\draw [color={rgb, 255:red, 74; green, 144; blue, 226 }  ,draw opacity=1 ][line width=1.5]    (170,89.36) -- (169.82,119.36) ;
		\draw [color={rgb, 255:red, 74; green, 144; blue, 226 }  ,draw opacity=1 ][line width=1.5]    (169,89.36) -- (458.82,90.18) ;
		\draw [color={rgb, 255:red, 74; green, 144; blue, 226 }  ,draw opacity=1 ][line width=1.5]    (459.82,89.18) -- (459.82,117.18) ;
		\draw [shift={(459.82,121.18)}, rotate = 270] [fill={rgb, 255:red, 74; green, 144; blue, 226 }  ,fill opacity=1 ][line width=0.08]  [draw opacity=0] (13.4,-6.43) -- (0,0) -- (13.4,6.44) -- (8.9,0) -- cycle    ;
		\draw  [fill={rgb, 255:red, 245; green, 166; blue, 35 }  ,fill opacity=1 ] (235.14,329.64) -- (247.53,324.74) -- (247.57,327.46) -- (487.23,327.46) -- (487.07,315) -- (484.36,315) -- (489.1,302.54) -- (494.16,315) -- (491.44,315) -- (491.65,331.83) -- (247.62,331.83) -- (247.66,334.54) -- cycle ;
		
		\draw (406,172) node [anchor=north west][inner sep=0.75pt]   [align=left] {Fuel Generators};
		\draw (301,181) node [anchor=north west][inner sep=0.75pt]   [align=left] {RESs};
		\draw (155,195) node [anchor=north west][inner sep=0.75pt]   [align=left] {EMS};
		\draw (332,283) node [anchor=north west][inner sep=0.75pt]   [align=left] {Loads};
		\draw (31,309) node [anchor=north west][inner sep=0.75pt]   [align=left] {Main Grid};
		\draw (203,357.64) node [anchor=north west][inner sep=0.75pt]   [align=left] {Grid Bus};
		\draw (489,357.82) node [anchor=north west][inner sep=0.75pt]   [align=left] {Microgrid};
		\draw (454.09,283.54) node [anchor=north west][inner sep=0.75pt]   [align=left] {Unified ESS};

	\end{tikzpicture}
	
	}
	\caption{\qt{Microgrid configuration with control actions (blue arrows) executed by the Energy Management System (EMS), including fuel generator output, load curtailment, and energy exchange with the main grid. Orange arrows denote bidirectional power flow (unified Energy Storage System (ESS) and main grid), red arrows indicate loads (consumption only), and green arrows represent Renewable Energy Sources (RESs) and fuel generators (supply only).}}
	\label{fig:mg_configuration}
\end{figure}
\section{Microgrid Modeling}
\label{sec:2-modeling}
\noindent \textbf{Notations:} \qt{The set of (non-negative) real numbers is denoted by $\bR$ ($\bR_+$), The set of (positive) natural numbers is denoted by $\bN$ ($\bN_+$), and $\bI_{[a:b]} := \bN \cap [a, b]$ for $0 \leq a \leq b$. The zero and one column vectors of length $n$ are denoted by $\bzero_n$ and $\bone_n$, respectively.} The operator $\|\cdot\|$ denotes the $2$-norm for vectors. Given $\cX \subset \bR^n$ and $\bbx \in \bR^n$, the projection operator $\Proj_{\cX}(\cdot)$ is defined by $\Proj_{\cX}(\bbx):=\min_{\bbx'\in\cX}\|\bbx - \bbx'\|$. \qt{The Cartesian product of sets $\cX_1$ and $\cX_2$ is denoted by $\cX_1 \times \cX_2$, and $\prod^n_{i=1}\cX_i := \cX_1 \times \cX_2 \times \cdots \times \cX_n$.} 
\\

This section provides the modeling of a typical grid-connected microgrid, consisting of fuel generators, an ESS, a unified RES, and curtailable loads~\citep{hu2023economic}. The considered model unifies the ones proposed by~\cite{hu2023economic} and ~\cite{pippia2019single} with slight modifications, and \textit{discrete-time} formulations are considered universally \qt{with the sampling interval and the time index denoted, respectively, by $\Ts$ and $t$.} A visualization of the considered microgrid configuration is given in Fig.~\ref{fig:mg_configuration}.

\subsection{Fuel Generators}
Consider $\Nfg$ independent fuel generators. For the $i$-th generator ($i \in \bI_{[1:\Nfg]}$), its operational cost is given by
\begin{multline}
	\label{eq:cost_generator}
	\Cfg{i}(t) = \Ts\theta_{1,i}\Pfg{i}(t) + \Ts^2\theta_{2,i}\Pfg{i}^2(t) + \Ts\Ofg{i}\dfg{i}(t) \\ + \Sfg{i}^{\mathrm{[on]}}\dfg{i}(t)(1 - \dfg{i}(t-1)) + \Sfg{i}^{\mathrm{[off]}}\dfg{i}(t-1)(1 - \dfg{i}(t)), 
\end{multline}
where $\Ts$ is the sampling interval, the binary variable $\dfg{i}(t) \in \{0, 1\}$ represents the OFF ($0$)/ON ($1$) mode of the generator, and $\Pfg{i}(t) \geq 0$ is the power generation. In \eqref{eq:cost_generator}, the polynomial $\Ts\theta_{1,i}\Pfg{i}(t) + \Ts^2\theta_{2,i}\Pfg{i}^2(t)$ is the fuel consumption cost, $\Ts\Ofg{i}\dfg{i}(t)$ is the operational cost, and $\Sfg{i}^{\mathrm{[on]}}$ and $\Sfg{i}^{\mathrm{[off]}}$ are the switch cost when starting up and shutting down the $i$-th generator, respectively. Besides, the following logic relation should be satisfied:
\begin{subequations}
	\label{eq:logic_generator}
	\begin{align}
		& \dfg{i}(t) = 0 \iff \Pfg{i}(t) = 0, \\
		& \dfg{i}(t) = 1 \iff \Pfg{i}(t) > 0.
	\end{align}
\end{subequations}
No dynamic behavior is considered for the fuel generators.

\subsection{Energy Storage Units}
\qt{Consider a unified ESS unit\footnote{The ESS is represented as a single aggregated unit (i.e., virtual battery), as commonly done in microgrid energy management to capture the overall energy balance and reflect centralized coordination of storage resources~\citep{parisio2014model, olivares2014trends}.}, whose State of Charge (SoC), denoted by $\xess(t)$,} is governed by the following \qt{\textit{piecewise-affine} (PWA) dynamics}~\citep{pippia2019single}:
\begin{equation}
	\label{eq:dynamics_ess}
	\qt{\xess(t+1) - \xess(t) = \fess(\Pess(t)) =
	\begin{cases}
		\Ts\etac\Pess(t) - \Ts \xdg & \Pess(t) \geq 0 \\
		\Ts\etad^{-1}\Pess(t) - \Ts \xdg & \Pess(t) < 0
	\end{cases}},
\end{equation}
\qt{where $\fess:\bR \to \bR_+$ is the PWA dynamics function, $\Pess(t) \in \bR$ is the ESS power flow, $\xdg$ is the constant energy degradation~\citep{hu2023economic}, and $\etac$ and $\etad$ are the charging and discharging coefficients, respectively. Battery energy storage systems exhibit round-trip efficiencies typically smaller than $1$, i.e., $\etac\etad < 1$~\citep{hittinger2015evaluating, pippia2019single}, depending on the technology and operating conditions, due to unavoidable electrochemical and resistive losses during charging and discharging. A binary variable $\dess(t) \in \{0,1\}$ is used to characterize the discharging ($0$)/charging ($1$) mode, and additional logic relations involving $\dess(t)$ and $\Pess(t)$ are given by}
\begin{subequations}
	\label{eq:logic_ess}
	\begin{align}
		& \dess(t) = 1 \iff \Pess(t) \geq 0, \\
		& \dess(t) = 0 \iff \Pess(t) < 0.
	\end{align}
\end{subequations}
The operational cost related to the $i$-th ESS unit is given by
\begin{equation}
	\label{eq:cost_ess}
	\qt{\Cess(t) = \Oess\Ts|\Pess(t)|,}
\end{equation}
where $\Oess$ is the operational cost coefficient of the ESS unit. 

\kk{\begin{remark}
	The unified ESS model considered in this work is sufficient to capture microgrid-level energy balancing~\citep{parisio2014model, pippia2019single, hu2023economic}, and also covers the case of multiple homogeneous ESS units~\citep{gao2021online}. Modeling heterogeneous ESS units primarily introduces device-level decisions (e.g., state-of-charge allocation and cycling), which are typically addressed within the low-level battery management systems~\citep{pozzi2020balancing}. In essence, the presence of multiple ESS units leads to hierarchical or bilevel control structures that couple grid-level scheduling with local storage coordination.
\end{remark}} 

\subsection{Renewable Energy Sources}
Following~\cite{hu2023economic}, we consider a \textit{unified} RES that aggregates multiple generation technologies, such as wind turbines and solar panels. The total generated power of the unified RES is denoted by $\Pres(t)$, which is constrained as
\begin{equation}
	\label{eq:cons_res}
	\Pres^- \leq \Pres(t) \leq \Pres^+.
\end{equation}
Although physically connected to the microgrid, the output power of the RES is uncontrollable and is therefore treated as an exogenous input. Moreover, its internal dynamics and operational costs are not taken into account for the microgrid optimization.

\subsection{Curtailable Loads}
In this paper, all loads are treated as a single aggregate that consumes power $\Pload(t)$. In addition, partial curtailment of the loads is allowed, while respecting the limits of users’ tolerance for discomfort~\citep{hu2023economic}. The curtailed load percentage is denoted by $\beta(t) \in [0,1]$, which is one of the \textit{controllable} inputs. The cost by supplying the loads is then given by
\begin{equation}
	\label{eq:cost_loads}
	\Cload(t) = \rho\beta(t)\Ts\Pload(t),
\end{equation}
where $\rho$ is the penalty weight on curtailments. Similar to $\Pres(t)$, the load power $\Pload(t)$ is bounded as
\begin{equation}
	\label{eq:cons_load}
	\Pload^- \leq \Pload(t) \leq \Pload^+,
\end{equation}
and it is also an \textit{exogenous} input to the system that cannot be managed by the microgrid itself.

\subsection{Power Exchange \& Power Balance}
The microgrid can trade electricity with the main grid to meet load demand efficiently and to enhance the overall economic benefit of operation. This exchanged power is denoted by $\Pexg(t) \in \bR$ and its associated binary variable $\dexg(t) \in \{0,1\}$ describes the selling ($0$)/purchasing ($1$) decision. The cost of power exchange is given by
\begin{equation}
	\label{eq:cost_exchange}
	\Cexg(t) = 
	\begin{cases}
		\cp(t)\Ts\Pexg(t) & \Pexg(t) \geq 0 \\
		\cs(t)\Ts\Pexg(t) & \Pexg(t) < 0 
	\end{cases},
\end{equation}
where $\cp$ and $\cs$ are the \qt{known \textit{time-varying} purchasing and selling prices}\footnote{\qt{The microgrid is considered to exchange power with the main grid through the day-ahead electricity market. Consequently, the electricity prices for each time period are known a priori, following the standard setting as in~\citep{vasilj2017day, alarcon2025learning}.}}, respectively. Besides, $\dexg(t)$ and $\Pexg(t)$ satisfy
\begin{subequations}
	\label{eq:logic_exchange}
	\begin{align}
		& \dexg(t) = 1 \iff \Pexg(t) \geq 0, \\
		& \dexg(t) = 0 \iff \Pexg(t) < 0.
	\end{align}
\end{subequations}
Finally, the powers of all the units in the microgrid must satisfy the following balance equation:
\begin{equation}
	\label{eq:balance}
	\sum^{\Nfg}_{i=1}\Pfg{i}(t) + \Pexg(t) + \Pres(t) = \Pess(t) + (1 - \beta(t))\Pload(t),
\end{equation}
where $(1 - \beta(t))\Pload(t)$ is the consumed power by the loads after curtailment.
\section{Economic Model Predictive Control for Microgrid Energy Management}
\qt{In this section, we formulate the EMPC objective by incorporating the operational costs introduced in Section~\ref{sec:2-modeling} and impose additional constraints on the control inputs and the state of charge (SoC) of the ESS unit. The resulting EMPC problem is then presented, together with a sufficient condition for recursive feasibility and a discussion of stability.}
\label{sec:3-mpc}
\subsection{\qt{Objective Function and Additional Constraints}}
\label{subsec3.1:obj_cons_empc}
The total operational cost of the microgrid is the sum of the individual costs of its components. \qt{These include the operational cost of the fuel generators~\eqref{eq:cost_generator}, the charging and discharging cost of the unified energy storage system (ESS)~\eqref{eq:cost_ess}, the cost from load curtailment~\eqref{eq:cost_loads}, and the cost of exchanging power with the main grid~\eqref{eq:cost_exchange}. Specifically, the grid-level cost at time step $t$ is given by:}
\begin{equation}
	\label{eq:cost_microgrid}
	\Cgrid(t) = \sum^{\Nfg}_{i=1}\Cfg{i}(t) + \Cess(t) + \Cload(t) + \Cexg(t).
\end{equation}
Several variables are subject to operational constraints. First, the discharging/charging power and SoC of ESS unit $i$ are both required to stay within a certain range to protect the ESS units, i.e.,
\begin{subequations}
	\label{eq:state_constraints}
	\begin{align}
		\label{eq:state_constraints_1}
		& |\Pess(t)| \leq \bPess \\
		\label{eq:state_constraints_2}
		& \xess^{-} \leq \xess(t) \leq \xess^{+}.
	\end{align}
\end{subequations}
\kk{Considering the dynamics~\eqref{eq:dynamics_ess} without the presence of ESS power flow (i.e., $\Pess(t) = 0$), the critical degradation horizon of the ESS unit is defined as 
\begin{equation}
	\label{eq:degrade_horizon}
	\Ness = \frac{\xess^+ - \xess^-}{\Ts\xdg},
\end{equation}
which characterizes the number of time steps required for the ESS to degrade from its maximum to minimum SoC. In this work, $\xdg$ is assumed to be sufficiently small such that $\Ness \gg 1$ and $\frac{\xdg}{\etac} \ll \bPess$, ensuring that degradation remains negligible relative to the charging capability.} Besides, if a fuel generator is ON, the produced power and its variation are both constrained, i.e.,
\begin{subequations}
	\label{eq:input_constraints_fg}
	\begin{align}
		\label{eq:input_constraints_fg_1}
		\Pfg{i}^{-}\dfg{i}(t) \leq \Pfg{i}(t) \leq \Pfg{i}^{+}\dfg{i}(t), \\
		\label{eq:input_constraints_fg_2}
		|\Pfg{i}(t) - \Pfg{i}(t-1)| \leq \Delta\Pfg{i}\dfg{i}(t).
	\end{align}
\end{subequations}
Moreover, only the case $\Pfg{i}^{-} \leq \Delta\Pfg{i}$ is considered such that a fuel generator is always allowed to start up at time step $t$ when it is OFF at $t-1$, but not vice versa. The power exchange should not exceed a given allowance, i.e.,
\begin{equation}
	\label{eq:input_constraint_grid}
	|\Pexg(t)| \leq \bPexg,
\end{equation}
and $\beta^+$ is an upper bound for the curtail percentage, i.e.,
\begin{equation}
	\label{eq:input_constraint_curtail}
	0 \leq \beta(t) \leq \beta^{+}.
\end{equation} 
The scalars used in the constraints \eqref{eq:cons_res}, \eqref{eq:cons_load}, and  \eqref{eq:state_constraints}--\eqref{eq:input_constraint_curtail} (i.e., $\Pres^-$, $\Pres^+$, $\Pload^-$, $\Pload^+$, $\bPess$, $\xess^-$, $\xess^+$, $\Pfg{i}^-$, $\Pfg{i}^+$, $\Delta \Pfg{i}$, $\bPexg$, and $\beta^+$) are all \textit{constant} and \textit{positive}. 

\subsection{\qt{Economic Model Predictive Control Formulation}}
\label{subsec3.3:empc}
EMPC is used to dynamically minimize the operational cost. The state and input of the system are defined, respectively, as 
\begin{subequations}
	\label{eq:state_input}
	\begin{align}
		\label{eq:state_definition}
		x(t) &:= \xess(t);\\
		\label{eq:input_definition}
		\bbu(t) &:= \qt{[\bbu_{\mathrm{p}}^\top(t), \beta(t)]^\top},
	\end{align}
\end{subequations}
\qt{where the power input $\bbu_{\mathrm{p}}(t) := [\bPfg^\top(t), \Pexg(t)]^\top$ with $\bPfg(t) = [\Pfg{1}(t), \Pfg{2}(t), \dots, \Pfg{\Nfg}(t)]^\top$. The state constraint set $\cX$ is given by $\cX = [\xess^{-}, \xess^{+}]$, and the input constraint set $\cU_{\mathrm{p}}(t)$ for the power input $\bbup$ is given by $\cU_{\mathrm{p}}(t) = \left(\prod^{\Nfg}_{i=1}\cU_{\text{fg},i}(t)\right) \times [-\bPexg, \bPexg]$, where $\cU_{\text{fg},i}(t) = ([\Pfg{i}^{-},\Pfg{i}^{+}]\cup\{0\})\cap [\Pfg{i}(t-1)-\Delta\Pfg{i}, \Pfg{i}(t-1)+\Delta\Pfg{i}]$. The RES power generation and load consumption form the disturbance as follows: 
\begin{equation}
	\label{eq:disturance_definition}
	\bbw(t) := [\Pres(t), \Pload(t)]^\top,
\end{equation}
and the disturbance set is given by $\cW = [\Pres^{-}, \Pres^{+}] \times [\Pload^{-}, \Pload^{+}]$. Following the state definition~\eqref{eq:state_definition}, the dynamics of the system state is given by
\begin{equation}
	\label{eq:dynamics_ess_catenate}
	x(t+1) = x(t) + \fess(\Pess(t)),
\end{equation}
where $\fess$ is defined in~\eqref{eq:dynamics_ess}. Based on the power balance equation~\eqref{eq:balance}, the ESS power flow is given by
\begin{align}
	\label{eq:ess_power_total_expression}
	\Pess(t) &= \sum^{\Nfg}_{i=1}\Pfg{i}(t) + \Pexg(t) + \underbrace{\Pres(t) - (1 - \beta(t))\Pload(t)}_{:=w_{\beta}(t,\beta(t))} \notag \\
	&=\bone^\top_{\Nfg+1}\bbu_{\mathrm{p}}(t) + w_{\beta}(t,\beta(t)).
\end{align}
The ESS power flow expression~\eqref{eq:ess_power_total_expression} provides an important insight into the considered EMS: the power input $\bbup(t)$ and the \textit{curtailment-shaped virtual disturbance} $w_{\beta}(t,\beta(t))$ jointly influence the ESS power flow. Combining \eqref{eq:dynamics_ess_catenate} and \eqref{eq:ess_power_total_expression}, the state dynamics function is given by
\begin{equation}
	\label{eq:final_state_dynamics}
	x(t+1) = x(t) + \fess\left(\bone^\top_{\Nfg+1}\bbup(t) + w_{\beta}(t,\beta(t))\right).
\end{equation}
In addition, the virtual disturbance $w_{\beta}(t,\beta(t))$ satisfies $w_{\beta}(t,\beta(t)) \in [w^{-}_{\beta}, w^{+}_{\beta}]$, where the bounds $w^{-}_{\beta}$ and $w^{+}_{\beta}$ can be explicitly computed as follows:}
\begin{subequations}
	\label{eq:virtual_disturbance_bound}
	\begin{align}
		& \qt{w^{-}_{\beta} = \Pres^- - \Pload^+;} \\
		& \qt{w^{+}_{\beta} = \Pres^+ - (1-\beta^+)\Pload^{-}.}
	\end{align}
\end{subequations}
\qt{At each time step $t$, after receiving the true renewable power generation $\Prest(t-1)$ and the true load consumption $\Ploadt(t-1)$, which jointly form the true virtual disturbance as $w^\ast_{\beta}(t-1,\beta(t-1)) = \Prest(t) - (1 - \beta(t))\Ploadt(t)$, the true state (SoC of ESS unit) $x^\ast(t) = \xesst(t)$ is then determined.} Furthermore, in this work, a load forecaster~\citep{burg2021comparative} and a RES forecaster~\citep{iheanetu2022solar, piotrowski2022evaluation} are available, which provide the predicted disturbance $\hat{\bbw}(t) := [\hPres(t), \hPload(t)]^\top$ at time step $t$, where $\hPres(t)$ and $\hPload(t)$ are the predicted RES generation and load consumption, respectively. Similarly, the predicted virtual disturbance is defined as $\hat{w}_{\beta}(t,\beta(t)) := \hPres(t) - (1 - \beta(t))\hPload(t)$. 

\kk{To satisfy state constraint under possible forecast error, a tightened \textit{polytopic} input constraint set is designed for $\bbu_{\mathrm{p}}(t)$ as follows:
\begin{equation}
	\label{eq:tightened_power_input}
	\overline{\cU}_{\mathrm{p}}(t,x(t)) = \left\{\bbup \in \cU_{\mathrm{p}}(t) \middle| 
	\underbrace{\begin{bmatrix}
		\bone^\top_{\Nfg+1} \\
		-\bone^\top_{\Nfg+1} \\
	\end{bmatrix}}_{:=G}\bbup
	\leq
	\underbrace{\begin{bmatrix}
		\frac{s^+_{\alpha}(x(t)) + \Ts\xdg}{\Ts\etac} - w^+_{\beta}\\
		-\frac{1}{\Ts}\max\left\{\frac{s^-_{\alpha}(x(t)) + \Ts\xdg}{\etac}, \frac{s^-_{\alpha}(x(t)) + \Ts\xdg}{\etad^{-1}}\right\} + w^{-}_{\beta}
	\end{bmatrix}}_{:=h(x(t))}
	\right\},
\end{equation}
where $\alpha \in (0, 1)$ is a constant, $s^+_{\alpha}(x(t)) := (1-\alpha)(\xess^+ - x(t))$, $s^-_{\alpha}(x(t)) := (1-\alpha)(\xess^- - x(t))$. The considered EMPC formulation relies solely on the economic stage cost and does not include additional terminal ingredients (i.e., terminal constraints or a terminal cost). Consequently, formally guaranteeing feasibility of the EMPC problem under the tightened input-constraint set $\overline{\cU}_{\mathrm{p}}(t,x(t))$ and recursive feasibility of the proposed EMPC strategy requires additional conditions. In this work, conditions on the \textit{maximum} sampling interval $\Ts$ and the \textit{minimum} power exchange allowance $\bPexg$ are imposed as follows (see also Theorem~\ref{thm:feasibility_recursive} below):
\begin{subequations}
	\label{eq:non-empty-of-set}
	\begin{align}
		\label{eq:non-empty-of-set-1}
		\Ts &\leq \frac{\etad(1 - \alpha)(\xess^+ - \xess^{-})}{(w^+_{\beta} - w^{-}_{\beta})}; \\
		\label{eq:non-empty-of-set-2}
		\bPexg &\geq \max\bigg\{- w^{-}_{\beta} + \frac{\xdg}{\etac}, w^{+}_{\beta} + \Pfgsys - \frac{\xdg}{\etac} \bigg\},
	\end{align}
\end{subequations}
where $\Pfgsys := \sum^{\Nfg}_{i=1}\Pfg{i}^+$, $\alpha$ is defined as in~\eqref{eq:tightened_power_input}, and $\wbm$ and $\wbp$ are given in~\eqref{eq:virtual_disturbance_bound}. The condition~\eqref{eq:non-empty-of-set-1} can be satisfied by selecting a sufficiently small sampling interval. Alternatively, it can be ensured by aggregating homogeneous ESS units into a large virtual ESS, thereby making the range $\xess^+ - \xess^-$ sufficiently large. Meanwhile, the condition~\eqref{eq:non-empty-of-set-2} is typically satisfied in practice, since the value of $\bPexg$ is generally large in real-world applications~\citep{pippia2019single}.} Given a prediction horizon $T \in \bN_+$, the EMPC controller seeks to minimize the $T$-step-ahead cumulative cost subject to: \qt{the dynamics constraint \eqref{eq:final_state_dynamics}, the state constraints~\eqref{eq:state_constraints_2}, and the input constraints~\eqref{eq:input_constraint_curtail} and~\eqref{eq:tightened_power_input}}. \kk{Note that it is sufficient to impose the tightened input constraint only at the current time step $t$ to reduce conservatism since only the first input is applied in closed-loop operation.} As such, the EMPC optimization problem can then be formulated as
\begin{align*}
	\hspace*{-17ex}\pmpc(I_{t}(T)): & \min \sum^{t + T-1}_{\tau = t}\Cgrid(\tau)
\end{align*}
\vspace*{-4ex}
\begin{align*}
	\hspace*{8ex}\text{s.t.}
	& \; \eqref{eq:final_state_dynamics}, \forall \tau \in \bI_{[t:t+T-1]}; \\
	& \; x(\tau) \in \cX, \forall \tau \in \bI_{[t:t+T]}; \\
	& \; \bbu_{\mathrm{p}}(t) \in \overline{\cU}_{\mathrm{p}}(t,x(t)); \\
	& \; \bbup(\tau) \in \cU_{\mathrm{p}}(\tau), \forall \tau \in \bI_{[t+1:t+T-1]}; \\
	& \; \beta(t) \in [\beta^-, \beta^+], \forall \tau \in \bI_{[t:t+T-1]}; \\
	& \; x(t) = x^\ast(t); \\
	& \; \bbw(\tau) = \hat{\bbw}(\tau), \forall \tau \in \bI_{[t:t+T-1]}.
\end{align*}
The optimization problem $\pmpc(I_{t}(T))$ \qt{can be reformulated as a mixed-integer quadratic program (MIQP) after introducing auxiliary variables to handle logic constraints, PWA functions, and other bilinear terms. The details of the reformulation are given in~\ref{app1:miqp}.} The problem $\pmpc(I_{t}(T))$ is parameterized by the information tuple $I_{t}(T)$ defined as 
\begin{equation}
	\label{eq:information}
	I_{t}(T) := \big(x^\ast(t), \bPfg(t-1), \{\hat{\bbw}(\tau)\}^{t+T-1}_{\tau=t}, \qt{\{\cp(\tau), \cs(\tau)\}^{t+T-1}_{\tau=t}}\big),
\end{equation}
\qt{which consists of the true state $x^\ast(t)$ at the current time step $t$, the power input $\bPfg(t-1)$ at the previous time step $t-1$, the predicted disturbance $\{\hat{\bbw}(\tau)\}^{t+T-1}_{\tau=t}$, and the price profile $\{\cp(\tau), \cs(\tau)\}^{t+T-1}_{\tau=t}$.} For simplicity, all generators are considered to be OFF initially, i.e., $\Pfg{i}(-1) = 0, \forall i \in \bI_{[1:\Nfg]}$. Solving $\pmpc(I_{t}(T))$ returns $\{\bbu^\star(\tau;\qt{I_{t}(T)})\}^{t+T-1}_{\tau=t}$, and only $\bbu^\star(t;I_{t}(T))$ is applied due to the moving-horizon mechanism of EMPC. Moreover, $\bbu^\star(t;I_{t}(T))$ implicitly defines an EMPC control policy as $\pimpc(I_{t}(T)) := \bbu^\star(\cdot;I_{t}(T))$ \citep{rawlings2017model}. After obtaining the true renewable power generation $\Prest(t)$ and the true load consumption $\Ploadt(t)$, and applying $\pimpc(I_{t}(T))$, the system state evolves to $x^\ast(t+1)$, and then the EMPC optimization is solved again. The optimization-based EMPC policy $\pi_{\mathrm{MPC}}$ is hereafter referred to as the expert EMPC policy. \qt{The following theorem provides the feasibility and recursive feasibility guarantee of the EMPC controller.
\begin{theorem}[Recursive Feasibility]
	\label{thm:feasibility_recursive}
	Let conditions \eqref{eq:non-empty-of-set-1} and \eqref{eq:non-empty-of-set-2} hold. Then, at time step $t$, given $x(t) \in \cX$, the set $\overline{\cU}_{\mathrm{p}}(t,x(t))$ is nonempty for all $x(t) \in \cX$, and the problem $\pmpc(I_{t}(T))$ is thus feasible. Moreover, the problem $\pmpc(I_{t+1}(T))$ is feasible for any possible disturbance prediction $\{\hat{\bbw}(\tau)\}^{t+T}_{\tau=t+1} \in \cW^{T}$ at the next time step $t+1$, i.e., the EMPC optimization problem is recursively feasible.  
\end{theorem}
The proof of Theorem~\ref{thm:feasibility_recursive} is given in~\ref{app2:proof_thm1}, and it essentially establishes that $\cX$ is a robust control invariant set based on \eqref{eq:non-empty-of-set-2}, and such an input that respects the input constraint and renders $\cX$ forward invariant can always satisfy the power balance condition.

\begin{remark}[Stability of EMPC for Energy Management]
	Unlike classical MPC for stabilization or reference tracking~\citep{rawlings2017model}, where stability of the closed‑loop system can be established via Lyapunov arguments owing to the positive-definite stage cost, standard EMPC does not inherently guarantee asymptotic stability because the economic stage cost is not necessarily positive definite and does not enforce convergence to a fixed setpoint~\citep{ellis2014tutorial}. Stability in EMPC typically requires additional conditions such as dissipativity, terminal costs, or terminal constraints~\citep{lin2023self}, which are not part of the basic formulation for EMPC in energy management applications~\citep{hu2023economic}. In essence, the primary objective of EMPC is to minimize the economic operational cost of the microgrid, instead of stabilizing the system state at a (time-invariant) equilibrium.
\end{remark}}

\kk{\begin{remark}[Feasibility and Microgrid Design]
	The feasibility condition~\eqref{eq:non-empty-of-set} can be interpreted as design guidelines for microgrid and its EMS that ensure robust energy management. In particular, these conditions impose requirements on key design parameters, including the sampling interval used for system operation, the capacity and operational limits of the ESS, and the interaction protocols with the main grid (e.g., energy purchasing/selling allowances). Satisfying these conditions is therefore closely tied to appropriate system sizing and control design choices, which jointly guarantee the existence of feasible energy management policies. Note that similar considerations have been widely recognized in the microgrid control and energy management literature~\citep{lasseter2011smart, pippia2019single,hu2023economic,alarcon2025artificial}.
\end{remark}
}
\section{Direct Approximate Mixed-Integer Model Predictive Control via Imitation Learning}
\label{sec:4-imitation}
This section discusses the details of IL-based approximate EMPC for microgrid energy management. In Section~\ref{subsec:4.1-motivation}, the motivations for directly approximating MI-MPC policy and using IL are further discussed. Section~\ref{subsec:4.2-imitation_learning} elaborates on the details of the proposed methodology, covering learning paradigm, feature design, data generation, and noise injection used to handle distribution shift.

\subsection{Motivations}
\label{subsec:4.1-motivation}
The EMPC optimization problem $\pmpc(I_{t}(T))$ exhibits several notable characteristics. First, the system has only $\Nfg+2$ control inputs, while the number of integer variables required to formulate $\pmpc(I_{t}(T))$ equals $2T(\Nfg+1)$, including $\dfg{i}(\tau)$, $\dess(\tau)$, $\dexg(\tau)$, and $\zfg{i}(\tau)$, for $\tau \in \bI_{t:t+T-1}$. It is obvious that $2T(\Nfg+1)> \Nfg+2$, thus the number of outputs of the neural network used to approximate the EMPC policy is smaller when adopting a direct approximate EMPC approach than when using indirect approaches that aim to learn integer variables instead~\citep{gao2021online, dasilva2025integrating}, especially for long prediction horizons. Moreover, among the total $2T(\Nfg+1)$ integer variables, $2(T-1)(\Nfg+1)$ (corresponding to a fraction $1-1/T$ of the total) are directly coupled with the predicted future inputs from $\tau = t+1$ to $\tau = t+T-1$, which are of limited interest since the primary objective is to learn the control policy only at the current step $t$. 

Most importantly, the power balance constraint~\eqref{eq:balance} induces a strong coupling between the control inputs $u(t)$ and the disturbances $w(t)$. Consequently, the integer variables also become \textit{disturbance-dependent}, as they are partially determined by the inputs through the logical constraints (see \eqref{eq:logic_generator} and \eqref{eq:logic_exchange}). The optimality of the learned integer variables thus relies heavily on the accuracy of the RES and load forecasts. In practice, forecast errors tend to grow with the prediction horizon, especially for renewable generation~\citep{iheanetu2022solar, piotrowski2022evaluation} and load demand~\citep{burg2021comparative} in microgrids. Longer-term forecasts are therefore more uncertain, which can result in higher operational costs and increased variability in system performance. As a result, indirect methods, which heavily rely on the predicted future disturbances to determine integer variables, are more sensitive to these errors and less robust. In contrast, direct methods, which only determine the current control action, are inherently less affected by such inaccuracies since the plant-model mismatch in MPC have exponential-decaying impact on the current input~\citep{lin2021perturbation, liu2024stability, liu2026certainty}. In short, for microgrid energy management, direct approximate EMPC constitutes a more parsimonious and robust approach, as it reduces output dimensionality and limits the influence of forecast uncertainty on policy optimality.

\qt{On the other hand, due to the presence of the generator switching costs~\eqref{eq:cost_generator}, the input constraint~\eqref{eq:input_constraints_fg_2}, the exogenous disturbances, and the price profile, the number of parameters required to characterize the parametric EMPC policy $\pimpc$ amounts to $1+\Nfg+4T$.} This high dimensionality makes the conventional approximate MPC with grid-based sampling~\citep{chen2018approximating, hertneck2018learning} computationally infeasible, particularly for long prediction horizons or systems with an increased number of fuel generators. Consequently, sampling state-input data from \textit{closed-loop} trajectories provides a more scalable alternative~\citep{drgovna2018approximate, karg2018deep}, which falls within the broader class of IL-based approximate MPC methods~\qt{\citep{pfrommer2024sample, pozzi2025imitation}}.

\subsection{Imitation Learning for Approximate EMPC}
\label{subsec:4.2-imitation_learning}
\qt{Since the inputs~\eqref{eq:input_definition} are continuous-valued, approximating $\pi_{\mathrm{MPC}}$ can be formulated as a regression problem. Consequently, supervised learning with a standard MLP is sufficient to obtain an accurate approximate controller~\citep{karg2018deep, pfrommer2024sample}. In this context, our imitation learning approach naturally reduces to behavior cloning~\citep{ross2011reduction, laskey2017dart}, in which a neural network policy is trained to directly mimic the expert EMPC policy $\pi_{\mathrm{MPC}}$. The focus of this paper is not on applying alternative neural network architectures to approximate MPC or on developing novel network structures.}

\subsubsection{Feature Design}
\qt{The state dynamics \eqref{eq:final_state_dynamics} indicates that the virtual disturbance $\hat{w}_{\beta}(t, \beta(t))$, which depends on $\beta(t)$, is affecting the predicted state trajectory when solving $\pmpc(I_t(T))$.} Therefore, an extra feature $\hat{\bw}_{\beta}(t)$ is designed as
\begin{equation}
	\label{eq:heuristic_feature}
	\hat{\bw}_{\beta}(t) = [\hat{w}_{\beta}(t, \beta_1), \hat{w}_{\beta}(t, \beta_2), \dots, \hat{w}_{\beta}(t, \beta_{N_\beta})]^\top,
\end{equation}
where $N_\beta$ is the resolution of $\hat{\bw}_{\beta}(t)$, and $\beta_i := \frac{(i-1)\beta^+}{N_{\beta}-1}$ for $i \in \bI_{[1:N_\beta]}$ with $\beta^+$ given in \eqref{eq:input_constraint_curtail}. Incorporating $\hat{\bw}_{\beta}(t)$, an augmented information tuple $\bar{I}_t(T)$ is defined by
\begin{equation}
	\label{eq:augmented_tuple}
	\bar{I}_t(T) := \left(I_t(T), \{\hat{\bw}_{\beta}(\tau)\}^{t+T_{w}-1}_{\tau = t}\right),
\end{equation}
where $I_t(T)$ is given in \eqref{eq:information} and $T_{w} \in \bI_{[0:T]}$ is the horizon depth of this extra feature. While increasing the resolution and horizon depth can enrich the feature set, it also leads to longer offline training and may necessitate a larger network with additional neurons in the hidden layers.

\subsubsection{Data Generation With Noise Injection}
Given a control horizon $\Tsim$, define a scenario $S$ as a tuple 
\begin{equation*}
	S := \left(x(0), \left\{\left(\Ploadt(t), \Prest(t)\right)\right\}^{\Tsim-1}_{t=0}\right).
\end{equation*}
For $\Nsim$ simulation scenarios, the training data set $\cD$ is obtained by rolling out \eqref{eq:final_state_dynamics} $\Tsim$ times for each of the scenarios. However, it is well known that behavior cloning suffers from distribution shift~\citep{ross2011reduction, laskey2017dart}, which degrades its performance. Therefore, inspired by~\cite{laskey2017dart}, we apply a simple off-policy noise injection technique, which can be effective in continuous control and computationally cheaper than on-policy methods (e.g., Dagger~\qt{\citep{ross2011reduction, pozzi2023imitation}}) to tackle distribution shift. Specifically, given a convariance matrix $\Sigma \in \bR^{(\Nfg+2)\times(\Nfg+2)}$, the noisy expert input for the $j$-th scenario is given by
\begin{equation}
	\label{eq:noisy}
	\umpct{t}^{[j]} = \pimpc\left(I^{[j]}_t(T)\right) + \epsilon^{[j]}_t,
\end{equation}
where $\epsilon^{[j]}_t \sim \cN(0, \Sigma)$ is Gaussian noise, and $I^{[j]}_t(T)$ is the information tuple acquired when simulating the $j$-th scenario. To satisfy the state and input constraints, post processing of the noisy input is needed. The final applied input $\mmpct{t}^{[j]}$ is obtained via
\begin{equation}
	\label{eq:projection}
	\mmpct{t}^{[j]} = \Proj_{\kk{\overline{\cU}_{\mathrm{p}}(t,x^\ast(t)) \times [\beta^-, \beta^+]}}\left(\umpct{t}^{[j]}\right),
\end{equation}
\kk{where $\overline{\cU}_{\mathrm{p}}(t,x^\ast(t))$ is defined in \eqref{eq:tightened_power_input}. Given, $x^\ast(t) \in \cX$ and $\mmpct{t}^{[j]} \in \cU_{\mathrm{p}}(t,x(t)) \times [\beta^-, \beta^+]$, it is guaranteed that $x^\ast(t+1) \in \cX$ (see more details in~\ref{app2:proof_thm1}).} As a result, the closed-loop system is generated by applying $\mmpct{t}^{[j]}$, and the training data set is given by
\begin{equation}
	\label{eq:data_set}
	\cD = \left\{\left(\bar{I}^{[j]}_t(T), \mmpct{t}^{[j]}\right)\right\}_{t=0,1,\dots,\Tsim, j=1,2,\dots,\Nsim},
\end{equation}
and the cardinality of $\cD$ is $|\cD| = \Tsim\Nsim$. 

\subsubsection{Loss Function, Training, and Post Processing} The learned policy using MLP is denoted by $\pimlp{\theta}$, which is parameterized by $\theta$, and it is a mapping from the space of augmented information tuples \qt{$\bar{\cI} \subset \bR^{1+\Nfg+4T+N_{\beta}T_{w}}$} to the input space $\cU \subset \bR^{\Nfg+2}$. The parameter $\theta$ includes all the weights and biases of the MLP, whose structure is depicted in Fig.~\ref{fig:mlp_network}.

\input{figure/mlp}

For the considered regression-based behavior cloning, the mean-squared-error loss is adopted as the learning metric~\citep{karg2018deep, pfrommer2024sample}, i.e., 
\begin{equation}
	\label{eq:final_loss}
	\hspace*{-2ex}
	\cL(\theta; \cD) := \sum^{\Nsim}_{j=1}\sum^{\Tsim-1}_{t=0}\left\| \pimlp{\theta}\left(\bar{I}^{[j]}_t(T)\right) - \mmpct{t}^{[j]}\right\|^2.\hspace*{-2ex}
\end{equation}
The training objective is to find the best policy $\pimlp{\theta^\star}$ parameterized by $\theta^\star$ through solving the following optimization problem:
\begin{equation}
	\label{eq:solving_theta}
	\theta^\star = \arg\min_{\theta}\cL(\theta;\cD). 
\end{equation}
In practice, this problem \eqref{eq:solving_theta} is typically non-convex and highly nonlinear, making it infeasible to guarantee convergence to the global optimum. Consequently, stochastic gradient descent or its variants are commonly used to find a local minimum $\hat{\theta}^\star$, which is then adopted as a suboptimal surrogate in most applications~\citep{goodfellow2016deep}. To guarantee that the learned policy satisfies \qt{the state and input constraints}, post processing of the MLP network output is needed. The final applied learning-based approximate EMPC input is given by\footnote{\qt{Since $\cU_{\mathrm{p}}(t,x^\ast(t)) \times [\beta^-, \beta^+]$ is a polytope, the projection in~\eqref{eq:projection} and~\eqref{eq:final_policy} is a convex optimization problem, which can be solved efficiently.}}
\begin{equation}
	\label{eq:final_policy}
	u_{\text{MLP},t,\hat{\theta}^\star} = \Proj_{\kk{\overline{\cU}_{\mathrm{p}}(t,x^\ast(t)) \times [\beta^-, \beta^+]}}\left(\pimlp{\hat{\theta}^\star}\left(\bar{I}_t(T)\right)\right),
\end{equation}
\kk{where $\overline{\cU}_{\mathrm{p}}(t,x^\ast(t))$ is defined in \eqref{eq:tightened_power_input}. Likewise, the input $u_{\text{MLP},t,\hat{\theta}^\star}$ can also guarantee constraint satisfaction due to the additional projection operation.}

\kk{\begin{remark}[Stochastic Extension]
	Recent studies have explored learning-based approaches to approximate stochastic MPC, demonstrating the potential of neural network-based approximations to reduce online computation in stochastic optimization~\citep{pozzi2025imitation}. In the current paper, we focus on certainty-equivalence MPC~\citep{meadows1995topics, liu2026certainty} with robust constraint tightening and its associated direct approximation strategy for microgrid energy management, where forecasts of loads and renewable generation are available, and the main challenge is learning a hybrid or mixed-integer control policy. Our approach emphasizes handling distribution shift and designing informative features to improve the accuracy and generalization of the learned controller. While the proposed framework is compatible with stochastic formulations, we adopt deterministic forecasts to simplify the control problem and clearly illustrate the efficacy of imitation learning in approximating MI-MPC laws. Extensions to explicitly handle stochastic forecast uncertainty in stochastic MPC are possible within the proposed framework and are left for future work.
\end{remark}

\begin{remark}[Constraint Satisfaction]
	\label{rmk:projection and state constraint}
	The projection operations in~\eqref{eq:projection} and~\eqref{eq:final_policy} ensure satisfaction of the state and input constraints when the input is perturbed due to injected noise or learning errors. This approach is typical to ensure constraint satisfaction (i.e., safety) when applying neural network-based controllers, similar to the methods of incorporating discrete-time control barrier functions~\citep{agrawal2017discrete, liu2025robust} to construct a safety filter that minimally modifies the MLP-based control input~\citep{cosner2022end}.
\end{remark}}
\section{Case Study}
\label{sec:5-simulation}
In this section, we present a \kk{numerical case study} of the proposed IL-based approximate EMPC approach, applied to a \kk{medium-scale} microgrid comprising photovoltaic panels, wind turbines, \kk{a unified ESS unit, and up to five fuel generators}. The case study considers a 24-hour time horizon (i.e., day-ahead scheduling) with a sampling interval of \kk{$\Ts = 5~\mathrm{min} = 1/12~\mathrm{h}$, resulting in a total of $\Tsim = 24 \cdot 12 = 288$ simulation steps.} All simulations are implemented in \qt{\texttt{Python 3.13.11}}, using \texttt{PyTorch} for the construction and training of the MLP network, while the MIQP problem of the expert EMPC is solved using \qt{\texttt{Gurobi 13.0.1}}~\citep{gurobi} with \texttt{gurobipy}.

\begin{figure}[h]
	\centering
	\includegraphics[width=\linewidth]{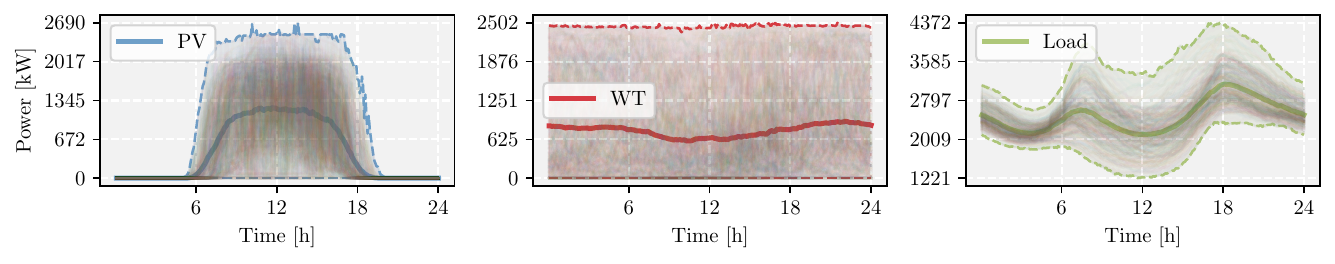}
	\caption{Visualization of \qt{$365$ disturbance realizations}, illustrating the power generated by photovoltaic (PV) panels and wind turbines (WT), as well as the load demand. \qt{The mean profile is depicted by a solid line, while the shaded region represents the range between the minimum and maximum values, which are indicated by dashed lines.}}
	\label{fig:profile}
\end{figure}

\kk{To account for diverse weather conditions affecting renewable energy generation (i.e., photovoltaic and wind power), load consumption, and electricity prices, we conduct a full-year simulation spanning $365$ days. The datasets for photovoltaic generation, wind power generation, load demand, and electricity prices are obtained from publicly available sources provided by the Australian Energy Market Operator, as also used by~\cite{huang2025grid}; a visualization of the datasets is given in Figure~\ref{fig:profile}.}

\begin{figure}[h]
	\centering
	\includegraphics[width=\linewidth]{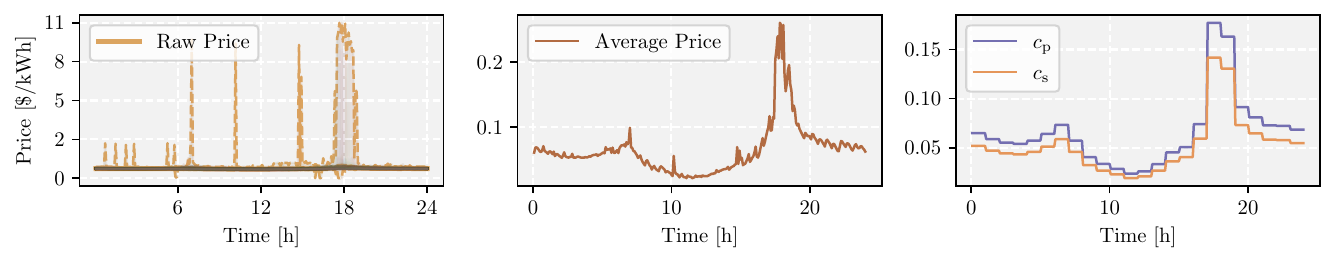}
	\caption{\qt{Visualization of (i) the raw electricity price data (left), including extreme outliers (the dashed lines are the min and max envolope), (ii) the corresponding average price after outlier removal (middle), and (iii) the resulting piecewise constant approximation of the purchasing and selling prices based on the average price (right).}}
	\label{fig:price}
\end{figure}

\kk{In this work, the electricity price is approximated by a piecewise constant profile constructed from the annual average of the raw price data, with extreme outliers removed (see Figure~\ref{fig:price}). This approximated price profile is used uniformly throughout the entire simulation horizon, and no price prediction error is considered. Theoretical analysis and numerical evaluation of performance degradation due to price prediction errors are beyond the scope of this paper and are left for future work. Furthermore, we adopt an asymmetric pricing scheme for grid interaction, where the selling price is set to $0.8$ times the purchasing price. Such a pricing structure, in which the selling price is lower than the purchasing price, is commonly adopted in the literature~\citep{pippia2019single, dasilva2025integrating}.}

\kk{The other used parameters and bounds of the considered microgrid is summarized in~\ref{app:table}. To account for seasonal variations, the disturbance realizations (i.e., photovoltaic generation, wind power generation, and load demand) from the first three weeks of each month, out of the total 365 days, are used for training and validation, while the remaining data are reserved for testing. In addition, four different initial states, $x(0) \in \{100\mathrm{kWh}, 500\mathrm{kWh}, 700\mathrm{kWh}, 900\mathrm{kWh}\}$, are considered, resulting in a total of $4 \cdot 21 \cdot 12 = 1008$ scenarios for training and validation. To simulate the effect of forecast inaccuracies, both the EMPC controller and the IL-based approximate EMPC controllers rely on a predicted disturbance signal corrupted by bounded noise uniformly distributed over $[-80\mathrm{kW}, 80\mathrm{kW}]$. At each time step, the noise is added to the true disturbance realization, and the resulting noisy disturbance realization is subsequently clipped at zero to ensure non-negativity. This approach aligns with standard practices where forecast errors are considered to be within known bounds~\citep{li2022data}.}
\begin{figure}[h]
	\centering
	\includegraphics[width=0.85\linewidth]{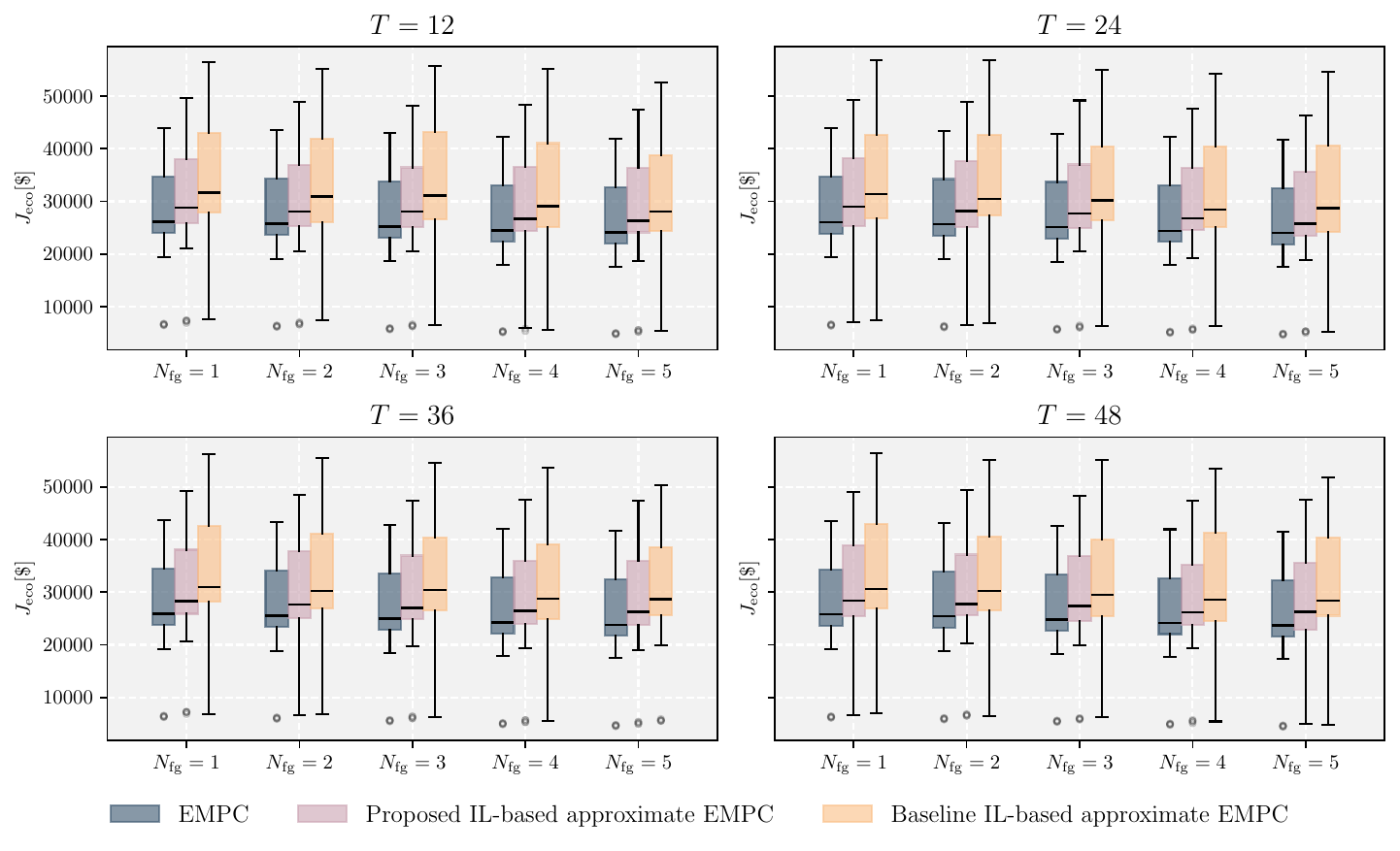}
	\caption{\qt{Comparison of the economic cost of three controllers when varying the prediction horizon $T$ and the number of fuel generators $\Nfg$: (a) the expert optimization-based economic model predictive control (EMPC) controller, (b) the proposed IL-based approximate EMPC controller, and (c) a baseline IL-based approximate EMPC controller without additional features and without the noise injection mechanism.}}
	\label{fig:scale_cost}
\end{figure}
\begin{figure}[h]
	\centering
	\includegraphics[width=0.85\linewidth]{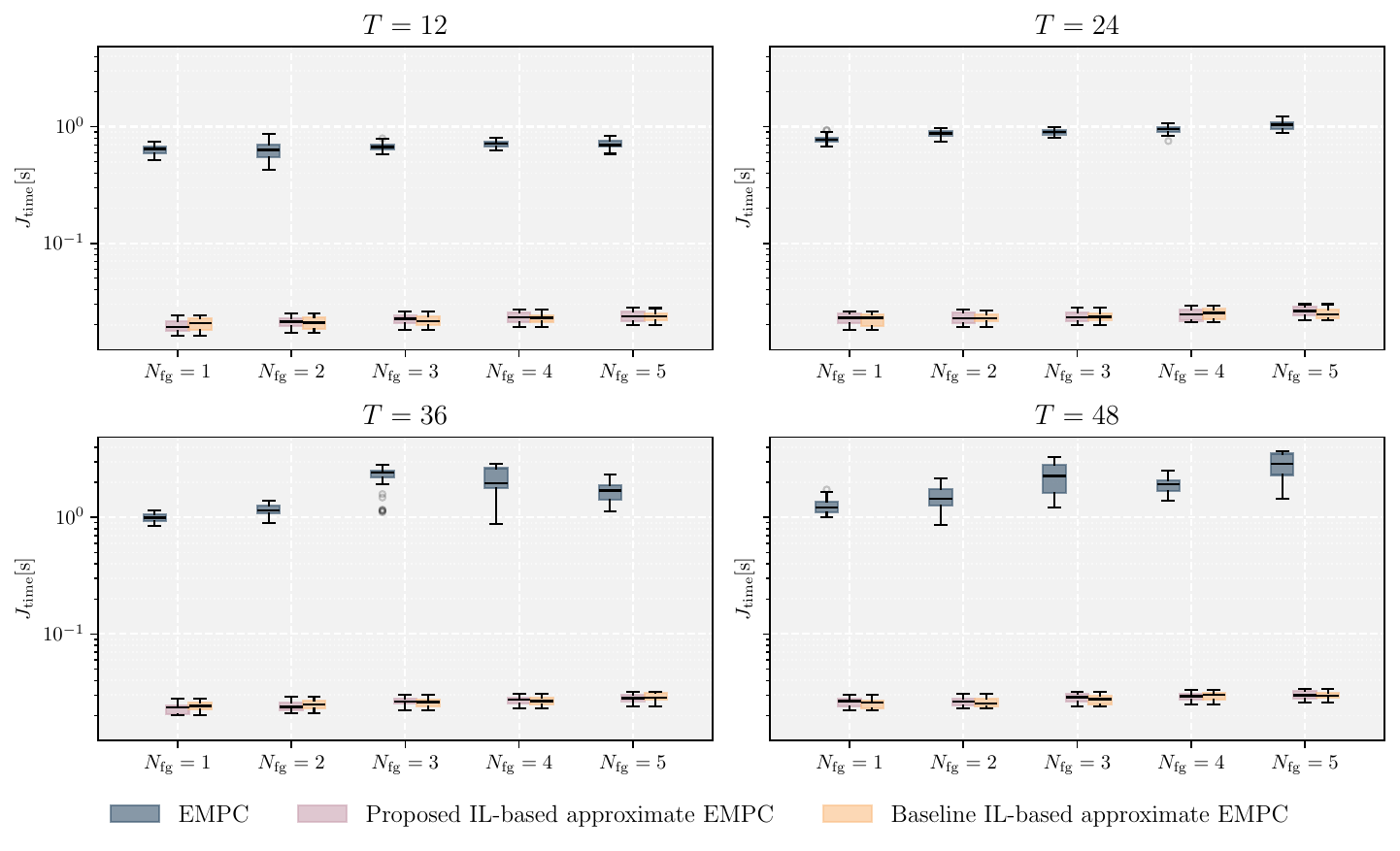}
	\caption{\qt{Comparison of the computation time of three controllers when varying the prediction horizon $T$ and the number of fuel generators $\Nfg$: (a) the expert optimization-based economic model predictive control (EMPC) controller, (b) the proposed IL-based approximate EMPC controller, and (c) a baseline IL-based approximate EMPC controller without additional features and without the noise injection mechanism.}}
	\label{fig:scale_time}
\end{figure}
	
Three different controllers are considered for performance comparison on the remaining $4\cdot 113=452$ test scenarios: (a) the expert optimization-based EMPC controller, (b) the proposed IL-based approximate EMPC controller, and (c) a baseline IL-based approximate EMPC controller without additional features and without the noise injection mechanism. The performance is evaluated using two metrics: the closed-loop economic cost and the computation time. The economic cost is defined as
\begin{equation*}
	J_{\text{eco}} := \sum_{t=0}^{\Tsim-1} \Cgrid(t),
\end{equation*}
where $\Cgrid(t)$ is given in~\eqref{eq:cost_microgrid}, and the computation time is defined as
\begin{equation*}
	J_{\text{time}} := \sum_{t=0}^{\Tsim-1} \Delta(t),
\end{equation*}
where $\Delta(t)$ denotes the \qt{CPU time}\footnote{\kk{All computations were performed on a machine equipped with an Intel Core i9 CPU at 32 GB RAM. Reported CPU times correspond to wall-clock time measured using Python function \texttt{time.perf\_counter()} on a single core.}} required to compute the control input at time step $t$. \kk{For the IL-based approximate EMPC controllers, $\Delta(t)$ denotes the time required to process the output of the MLP network given an input. In contrast, for the expert EMPC controller, $\Delta(t)$ accounts solely for the CPU time required to solve the optimization problem, excluding the time needed to construct and formulate it, as this is negligible compared to the solve time. Furthermore, because the optimization problem is parametric, the model only needs to be constructed once to preserve its structure, after which different parameter values can be supplied for each subsequent solve.}

\begin{figure}[h]
	\centering
	\includegraphics[width=0.85\linewidth]{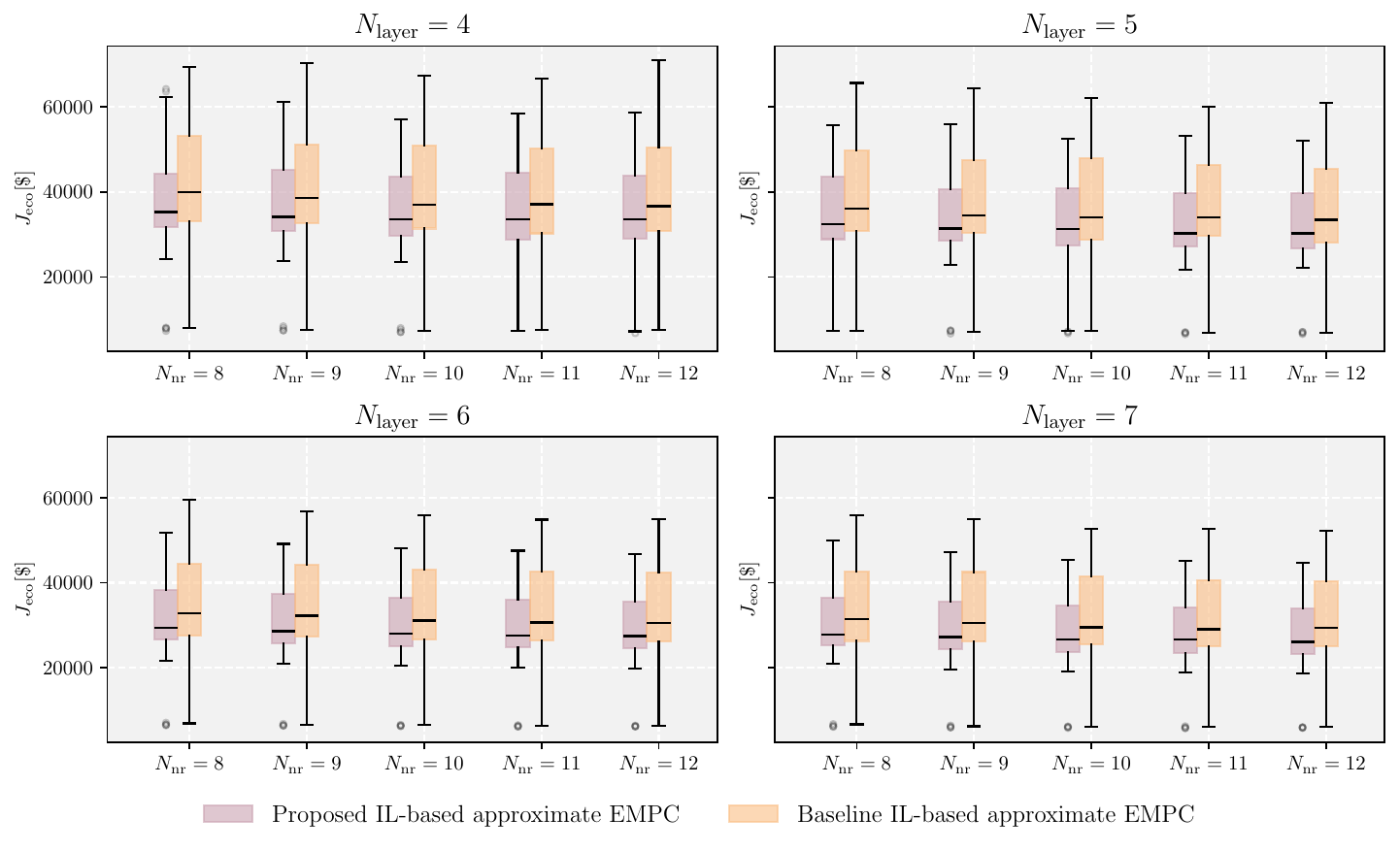}
	\caption{\qt{Comparison of the economic cost of two controllers when varying the number of layers $N_{\mathrm{layer}}$ and the number of neurons per layer $N_{\mathrm{nr}}$ of the multi-layer perceptron: (a) the proposed IL-based approximate economic model predictive control (EMPC) controller, and (b) a baseline IL-based approximate EMPC controller without additional features and without the noise injection mechanism.}}
	\label{fig:mlp_cost}
\end{figure}

\kk{The nominal configuration of the microgrid includes three fuel generators. For the EMPC controller, the nominal prediction horizon is set to $T = 12$, corresponding to a 1-hour ahead prediction window. To evaluate the scalability of the proposed approach and highlight the advantages of the neural network-based controller, we vary the number of active fuel generators as $\Nfg \in \{1,2,3,4,5\}$, and the EMPC prediction horizon as $T \in \{12,24,36,48\}$. The results for the economic cost and computation time are given, respectively, in Figure~\ref{fig:scale_cost} and Figure~\ref{fig:scale_time}. The figures clearly indicate that the proposed IL-based approach outperforms the baseline IL-based method across the evaluated scenarios, and that it has comparable performance with respect to the expert EMPC. Furthermore, both approximate MPC controllers achieve an approximately one-order-of-magnitude reduction in computation time compared to the expert MPC controller, demonstrating a substantial improvement in computational efficiency.}

\kk{The adopted MLP consists of $N_{\mathrm{layer}} = 6$ hidden layers, each with $N_{\mathrm{nr}} = 10$ neurons, and employs the GELU activation function~\citep{hendrycks2016gaussian}. Similar small-sized MLP networks have proved to be effective for approximate MPC~\citep{karg2018deep}. The network architecture is selected based on empirical evaluations on the nominal microgrid setting, considering multiple configurations with varying numbers of layers and neurons. For the extra feature related to the virtual disturbance $w_{\beta}(t, \beta(t))$, the hyperparameters of our proposed method are chosen as $N_\beta = 3$ and $T_w = 6$. We adopt an MLP as a standard function approximator due to its universal approximation properties and computational efficiency. The exact architecture is not critical to our approach, and Figure~\ref{fig:mlp_cost} shows consistent performance improvements of the proposed IL-based EMPC approach over the baseline IL-based EMPC approach across a range of reasonable configurations. While the number of layers and neurons influences the approximation capacity, the proposed method does not rely on a finely tuned architecture and remains applicable to alternative MLP configurations. Note that developing novel network architectures or exploring the best MLP configuration is not essential to the contribution of the current paper.}


\section{Conclusions and Future Work}
\label{sec:6-conclusion}
\qt{In this paper, an imitation learning (IL)–based framework is proposed for approximate EMPC applied to a microgrid energy management problem with curtailable loads. The proposed method outperforms a basic IL approach in achieving more optimal economic performance, leveraging a novel designed feature tailored for the disturbances and a noise injection mechanism to mitigate distribution shift. Besides, a novel input constraint-tightening approach is developed to ensure both satisfaction of both input and state constraints, achieving safe imitation learning. In simulation experiments, the learned approximate EMPC controller reduces the online computation time to approximately one order of magnitude lower than that of the optimization-based EMPC, while maintaining comparable economic performance.

Future work will focus on extending the proposed framework to stochastic EMPC for microgrids, explicitly accounting for distributional uncertainty in renewable generation and demand. The closed-loop performance under varying load consumption patterns will be systematically analyzed, and the EMPC control strategy itself will be further investigated. In addition, the impact of price prediction errors on economic performance will be studied through both theoretical and numerical analysis. Finally, the framework will be extended to incorporate heterogeneous storage units with low-level battery management systems and to address networked microgrids with inter-grid interactions.}

\appendix
\appendix
\section{\qt{Reformulation of EMPC Optimization Problem as Mixed-Integer Quadratic Program}}
\label{app1:miqp}
\qt{We first reformulate the basic logical constraints. Since \eqref{eq:input_constraints_fg_1} already provides a linear reformulation of \eqref{eq:logic_generator}, it remains to reformulate \eqref{eq:logic_ess} and \eqref{eq:logic_exchange} as follows:}
\begin{subequations}
	\label{eq:logic_reformulation_basic}
	\begin{align}
		\label{eq:logic_reformulation_ess}
		& \;\qt{-\bPess(1 - \dess(t)) \leq \Pess(t) \leq \bPess\dess(t),} \\
		\label{eq:logic_reformulation_exg}
		& \;\qt{-\bPexg(1 - \dexg(t)) \leq \Pexg(t) \leq \bPexg\dexg(t).}
	\end{align}
\end{subequations}
For optimization, the SoC dynamics~\eqref{eq:dynamics_ess}, ESS cost~\eqref{eq:cost_ess}, and exchange cost~\eqref{eq:cost_exchange} \qt{requires a reformulation due to their PWA nature}, respectively, as
\begin{subequations}
	\label{eq:reformulation_hybrid_initial}
	\begin{align}
		\label{eq:dynamics_ess_reformulated_initial}
		& \qt{\xess(t+1) = \xess(t) + \Ts(\etac - \etad^{-1})\dess(t)\Pess(t) + \Ts\etad^{-1}\Pess(t) - \Ts \xdg}, \\
		\label{eq:cost_ess_reformulated_initial}
		& \qt{\Cess(t) = \Oess\big(2\dess(t)\Pess(t) - \Pess(t)\big)}, \\
		\label{eq:cost_exchange_reformulated_initial}
		& \Cexg(t) = \qt{(\cp(t) - \cs(t))}\dexg(t)\Pexg(t) + \cs\Pexg(t).
	\end{align}
\end{subequations}
Accordingly, auxiliary \textit{continuous} variables $\zess(t) := \dess(t)\Pess(t)$ and $\zexg(t) := \dexg(t)\Pexg(t)$ are required to \qt{eliminate \textit{bilinear} terms} in the optimization formulation, i.e.,
\begin{subequations}
	\label{eq:reformulation_hybrid}
	\begin{align}
		\label{eq:dynamics_ess_reformulated}
		& \qt{\xess(t+1) = \xess(t) + \Ts(\etac - \etad^{-1})\zess(t) + \Ts\etad^{-1}\Pess(t) - \Ts \xdg}, \\
		\label{eq:cost_ess_reformulated}
		& \qt{\Cess(t) = \Oess\big(2\zess(t) - \Pess(t)\big)}, \\
		\label{eq:cost_exchange_reformulated}
		& \qt{\Cexg(t) = (\cp(t) - \cs(t))\zexg(t) + \cs\Pexg(t).}
	\end{align}
\end{subequations}
Moreover, additional linear constraints are needed \citep{bemporad1999control} to characterize $\zess(t)$ and $\zexg(t)$ as follows:
\begin{subequations}
	\label{eq:cons_auxiliary}
	\begin{align}
		\label{eq:cons_auxiliary_ess1}
		& -\dess(t)\bPess \leq \zess(t) \leq \dess(t)\bPess, \\
		\label{eq:cons_auxiliary_exg1}
		& -\dexg(t)\bPexg \leq \zexg(t) \leq \dexg(t)\bPexg, \\
		\label{eq:cons_auxiliary_ess2}
		& \Pess(t) - \bPess(1 - \dess(t)) \leq \zess(t), \\
		\label{eq:cons_auxiliary_ess3}
		& \zess(t) \leq \Pess(t) + \bPess(1 - \dess(t)), \\
		\label{eq:cons_auxiliary_exg2}
		& \Pexg(t) - \bPexg(1 - \dexg(t)) \leq \zexg(t), \\
		\label{eq:cons_auxiliary_exg3}
		& \zexg(t) \leq \Pexg(t) + \bPexg(1 - \dexg(t)).
	\end{align}
\end{subequations}
Besides, the bilinear term $\dfg{i}(t)\dfg{i}(t-1)$ in \eqref{eq:cost_generator} also needs to be reformulated by introducing an auxiliary \textit{binary} variable $\zfg{i}(t) := \dfg{i}(t)\dfg{i}(t-1)$ with the following additional linear constraints:
\begin{subequations}
	\label{eq:cons_abs_auxiliary}
	\begin{align}
		& \zfg{i}(t) \leq \dfg{i}(t), \\
		& \zfg{i}(t) \leq \dfg{i}(t-1), \\
		& \zfg{i}(t) \geq \dfg{i}(t-1) + \dfg{i}(t) - 1.
	\end{align}
\end{subequations}
In summary, the constraints of EMPC problem are: the dynamics constraint \eqref{eq:dynamics_ess_reformulated}; the power balance equation \eqref{eq:balance}, state and input constraints \eqref{eq:state_constraints}--\eqref{eq:input_constraint_curtail}; logic constraints \eqref{eq:logic_generator}, \eqref{eq:logic_ess}, and \eqref{eq:logic_exchange}; and other constraints \eqref{eq:cons_auxiliary} and \eqref{eq:cons_abs_auxiliary} involving the auxiliary variables $\zess(t)$, $\zexg(t)$ and $\zfg{i}(t)$. As such, the EMPC optimization problem can then be reformulated as the following MIQP:
\begin{align*}
	\hspace*{-17ex}\pmpc(I_{t}(T)): & \min \sum^{t + T-1}_{\tau = t}\Cgrid(\tau)
\end{align*}
\vspace*{-4ex}
\begin{align*}
	\hspace*{8ex}\text{s.t.}
	& \; x(\tau) \in \cX, \forall \tau \in \bI_{[t:t+T]}; \\
	& \; \bbup(\tau) \in \cU_{\mathrm{p}}(\tau), \forall \tau \in \bI_{[t+1:t+T-1]}; \\
	& \; \beta(t) \in [\beta^-, \beta^+], \forall \tau \in \bI_{[t:t+T-1]}; \\
	& \; \bbu_{\mathrm{p}}(t) \in \overline{\cU}_{\mathrm{p}}(t,x(t)); \\
	& \; \eqref{eq:logic_reformulation_exg}, \eqref{eq:cons_auxiliary_exg1}, \eqref{eq:cons_auxiliary_exg3}, \eqref{eq:cons_auxiliary_exg2}\; \forall \tau \in \bI_{[t:t+T-1]};  \\
	& \; \qt{\eqref{eq:logic_reformulation_ess}, \eqref{eq:dynamics_ess_reformulated}, \eqref{eq:cons_auxiliary_ess1}, \eqref{eq:cons_auxiliary_ess3}, \eqref{eq:cons_auxiliary_ess2}, \forall \tau \in \bI_{[t:t+T-1]}}; \\
	& \; \qt{\eqref{eq:cons_abs_auxiliary}, \forall i \in \bI_{[1:\Nfg]}, \tau \in \bI_{[t:t+T-1]};} \\
	& \; \qt{\eqref{eq:input_constraints_fg_1}, \forall i \in \bI_{[1:\Nfg]}, \tau \in \bI_{[t-1:t+T-1]};} \\
	& \; x(t) = x^\ast(t); \\
	& \; \bbw(\tau) = \hat{\bbw}(\tau), \forall \tau \in \bI_{[t:t+T-1]}.
\end{align*}

\qt{
\section{Proof of Theorem~\ref{thm:feasibility_recursive} (Recursive Feasibility of EMPC)}
\label{app2:proof_thm1}
The proof consists of three parts: (i) non-emptyness of the tightened input constraint set $\overline{\cU}_{\mathrm{p}}(t,x(t))$ for any feasible state $x(t) \in \cX$, (ii) feasibility of the true state $x^\ast(t+1)$ (i.e., $x^\ast(t+1) \in \cX$) after applying $\bbu(t) \in \overline{\cU}_{\mathrm{p}}(t,x(t))\times[\beta^-, \beta^+]$ under forecast error, and (iii) recursive feasibility of the EMPC optimization problem.
\begin{itemize}
	\item part (i): First, noting that $\etac \leq \etad^{-1}$, the condition~\eqref{eq:non-empty-of-set-1} implies that $\frac{s^+_{\alpha}(x(t))+\Ts\xdg}{\Ts\etac} - \frac{s^+_{\alpha}(x(t))+\Ts\xdg}{\Ts\etac} \geq \wbp-\wbm$ and $\frac{s^+_{\alpha}(x(t))+\Ts\xdg}{\Ts\etac} - \frac{s^+_{\alpha}(x(t))+\Ts\xdg}{\Ts\etad^{-1}} \geq \wbp-\wbm$, where $\alpha \in (0,1)$, $s^+_{\alpha}(x(t))$, and $s^-_{\alpha}(x(t))$ are given as in~\eqref{eq:tightened_power_input}, 
	meaning that $\{\bbu \in \bR^{\Nfg+1} \mid G\bbu \leq h(x(t))\} \neq \emptyset$. Moreover, the condition \eqref{eq:non-empty-of-set-2} leads to $\min_{\bbup(t-1)\in\cU_{\mathrm{p}}(t-1)}$ $\max_{\bbup(t)\in\cU_{\mathrm{p}}(t)}\bone^\top_{\Nfg+1}\bbup(t) = \bPexg + \sum_{i=1}^{\Nfg}\Delta\Pfg{i} \geq \frac{\xdg}{\etac} -\wbm =\max_{x(t)\in\cX}\frac{s^-_{\alpha}(x(t))}{\Ts\etac} -\wbm$ and $\max_{\bbup(t-1)\in\cU_{\mathrm{p}}(t-1)}\min_{\bbup(t)\in\cU_{\mathrm{p}}(t)}\bone^\top_{\Nfg+1}\bbup(t) = - \bPexg + \sum_{i=1}^{\Nfg}(\Pfg{i}^{+}-\Delta\Pfg{i}) \leq \frac{\xdg}{\etac} -\wbp = \min_{x(t)\in\cX}\frac{s^+_{\alpha}(x(t))}{\Ts\etac} -\wbp$. Moreover, noting $\etac \leq \etad^{-1}$, \eqref{eq:non-empty-of-set-2} also implies that $\min_{\bbup(t-1)\in\cU_{\mathrm{p}}(t-1)}$ $\max_{\bbup(t)\in\cU_{\mathrm{p}}(t)}\bone^\top_{\Nfg+1}\bbup(t) = \bPexg + \sum_{i=1}^{\Nfg}\Delta\Pfg{i} \geq \frac{\xdg}{\etad^{-1}} -\wbm =\max_{x(t)\in\cX}\frac{s^-_{\alpha}(x(t))}{\Ts\etad^{-1}} -\wbm$. Therefore, we have $\forall x(t) \in \cX$, $\cU_{\mathrm{p}}(t) \cap \{u \in \bR^{\Nfg+1} \mid Gu \leq h(x(t))\} \neq \emptyset$, i.e., $\forall x(t) \in \cX$, $\overline{\cU}_{\mathrm{p}}(t,x(t)) \neq \emptyset$.
	\item part (ii): Given $x^\ast(t) \in \cX$ and $\bbup(t) \in \overline{\cU}_{\mathrm{p}}(t,x^\ast(t))$, for all $\beta(t) \in [\beta^-, \beta^+]$, $G\bbup(t) \leq h(x^\ast(t))$ implies that $\Ts\etac(\bone^\top_{\Nfg+1}\bbup(t) + \wbb) - \Ts\xdg \geq (\alpha-1)[x^\ast(t) - \xess^{-}]$, $\Ts\etad^{-1}(\bone^\top_{\Nfg+1}\bbup(t) + \wbb) - \Ts\xdg \geq (\alpha-1)[x^\ast(t) - \xess^{-}]$, and $\Ts\etac(\bone^\top_{\Nfg+1}\bbup(t) + \wbb) - \Ts\xdg \leq (1-\alpha)[\xess^+ - x^\ast(t)]$. The above inequalities jointly lead to $[\xess^+ - x^\ast(t+1)] \geq \alpha [\xess^+ - x(t)]$ and $[x^\ast(t+1) - \xess^-] \geq \alpha [x^\ast(t) - \xess^-]$ for both $\Pess(t) \geq 0$ and $\Pess(t) < 0$. Therefore, $x^\ast(t+1) \in \cX$ under $\bbu(t) \in \overline{\cU}_{\mathrm{p}}(t,x^\ast(t)) \times [\beta^-, \beta^+]$.
	\item part (iii): Assume $\pmpc(I_{t}(T))$ is feasible for $x^\ast(t) \in \cX$. It is sufficient to prove the existence of one feasible solution to $\pmpc(I_{t+1}(T))$. From part (ii), it is known that $x^\ast(t+1) \in \cX$. Since the proof of part (ii) also holds for $\hat{w}_\beta(t,\beta(t))$, there exists $\bbup(t+1) \in \cU_{\mathrm{p}}(t+1,x^\ast(t+1))$ such that $x(t+2) \in \cX$. Condition~\eqref{eq:non-empty-of-set-2} implies that $\forall \hat{\bbw}(t) \in \cW$, there exists $\bbu(t) \in \cU_{\mathrm{p}} \times [\beta^-, \beta^+]$ such that $\bone^\top_{\Nfg+1}\bbup(t) + \hat{w}_\beta(t,\beta(t)) = \frac{\xdg}{\etac} \leq \bPess$, keeping $x(\tau) = x(t+2) \in \cX$ for all $\tau \in \bI_{t+3:t+T+1}$. Therefore, $\pmpc(I_{t+1}(T))$ is feasible.
\end{itemize}

\section{Microgrid Parameters}
\label{app:table}
\begin{table}[h]
	\label{tab:parameters}
	\centering
	\caption{\qt{Parameters and bounds used in the numerical simulation.}}
	{\renewcommand{\arraystretch}{0.95}
		\setlength{\tabcolsep}{2pt}
		\qt{
			\begin{tabularx}{0.92\columnwidth}{c c c}
				\toprule[0.3ex]
				Parameter & Value & Units \\
				\midrule
				$\theta_{1,i} (i=1,2,3,4,5)$ & [\num{4.8e-3}, \num{9.6e-3}, \num{7.2e-3}, \num{6.0e-3}, \num{8.4e-3}] & \si{\$/\kilo\watt\hour} \\
				
				$\theta_{2,i} (i=1,2,3,4,5)$ & [\num{8.64e-1}, \num{6.48e-1}, \num{1.73e-1}, \num{3.46e-1}, \num{5.45e-1}] & \si{\$/(\kilo\watt\hour)^{-2}} \\
				
				$\Ofg{i} (i=1,2,3,4,5)$ & [\num{7}, \num{12.5}, \num{9}, \num{10}, \num{11}] & \si{\$/\hour} \\
				
				$\Sfg{i} (i=1,2,3,4,5)$ & [\num{50}, \num{40}, \num{55}, \num{45}, \num{35}] & \si{\$} \\
				
				$\Pfg{i}^{-} (i=1,2,3,4,5)$ & [\num{200}, \num{210}, \num{230}, \num{220}, \num{190}] & \si{\kilo\watt\hour} \\
				
				$\Pfg{i}^{+} (i=1,2,3,4,5)$ & [\num{2.0e3}, \num{2.1e3}, \num{2.3e3}, \num{2.2e3}, \num{1.9e3}] & \si{\kilo\watt\hour} \\
				
				$\Delta\Pfg{i} (i=1,2,3,4,5)$ & [\num{300}, \num{315}, \num{345}, \num{330}, \num{285}] & \si{\kilo\watt\hour} \\
				
				$[\etac, \etad]$ & [\num{0.9}, \num{0.9}] & \si{--} \\
				
				$\xdg$ & \num{0.5} & \si{\kilo\watt} \\
				
				$\Oess$ & \num{0.005} & \si{\$/\kilo\watt\hour} \\
				
				$\bPess$ & \num{100} & \si{\kilo\watt} \\
				
				$[\xess^{-}, \xess^{+}]$ & [\num{120}, \num{1350}] & \si{\kilo\watt\hour} \\
				
				$\rho$ & \num{1.5} & \si{\$/\kilo\watt\hour} \\
				
				$[\Pload^{-}, \Pload^{+}]$ & [\num{1.2e3}, \num{4.4e3}] & \si{\kilo\watt} \\
				
				$\beta^+$ & \num{0.2} & \si{--} \\
				
				$\bPexg$ & \num{1.5e4} & \si{\kilo\watt} \\
				
				\bottomrule[0.3ex]
		\end{tabularx}}
	}
\end{table}
}

\bibliographystyle{elsarticle-harv} 
\bibliography{microgridref.bib}

\end{document}